\documentclass[longauth]{aa} 

\usepackage{color}

\usepackage{multirow}
\usepackage{changes}
\usepackage{orcidlink}
\usepackage[switch]{lineno} 

\usepackage[rightcaption]{sidecap}

\defcitealias{Nicolas_2021}{N21} 

\usepackage{graphicx}
\usepackage{txfonts}
\hypersetup{
    colorlinks=true,
    linkcolor=blue,
    filecolor=blue,      
    urlcolor=blue,
    citecolor=blue
    }

\begin{document}

   \title{ZTF SN Ia DR2: Environmental dependencies of stretch and luminosity of a volume limited sample of 1,000 Type Ia Supernovae}

   \author{
        Ginolin, M. \inst{1} \fnmsep\thanks{Corresponding author: \texttt{m.ginolin@ip2i.in2p3.fr}} \orcidlink{0009-0004-5311-9301} 
\and Rigault, M. \inst{1} \orcidlink{0000-0002-8121-2560}
\and Smith, M. \inst{1, 2} \orcidlink{0000-0002-3321-1432}
\and Copin, Y. \inst{1} \orcidlink{0000-0002-5317-7518} 
\and Ruppin, F.\inst{1} \orcidlink{0000-0002-0955-8954}
\and Dimitriadis, G. \inst{3} \orcidlink{0000-0001-9494-179X}
\and Goobar, A. \inst{4} \orcidlink{0000-0002-4163-4996}
\and Johansson, J. \inst{4} \orcidlink{0000-0001-5975-290X}
\and Maguire, K.\inst{3} \orcidlink{0000-0002-9770-3508}
\and Nordin, J. \inst{5} \orcidlink{0000-0001-8342-6274}
\and Amenouche, M. \inst{6} \orcidlink{0009-0006-7454-3579}
\and Aubert, M. \inst{7}
\and Barjou-Delayre, C. \inst{7}
\and Betoule, M. \inst{8} \orcidlink{0000-0003-0804-836X}
\and Burgaz, U.\inst{3} \orcidlink{0000-0003-0126-3999}
\and Carreres, B. \inst{9, 10} \orcidlink{0000-0002-7234-844X}
\and Deckers, M. \inst{3} \orcidlink{0000-0001-8857-9843}
\and Dhawan, S. \inst{11} \orcidlink{0000-0002-2376-6979}
\and Feinstein, F. \inst{9}
\and Fouchez, D. \inst{9} \orcidlink{0000-0002-7496-3796}
\and Galbany, L. \inst{12, 13} \orcidlink{0000-0002-1296-6887}
\and Ganot, C. \inst{1}
\and Harvey, L. \inst{3} \orcidlink{0000-0003-3393-9383}
\and de Jaeger, T. \inst{8} \orcidlink{0000-0001-6069-1139}
\and Kenworthy, W. D. \inst{4} \orcidlink{0000-0002-5153-5983}
\and Kim, Y.-L. \inst{2} \orcidlink{0000-0002-1031-0796}
\and Kowalski, M. \inst{5, 14} \orcidlink{0000-0001-8594-8666}
\and Kuhn, D. \inst{8} \orcidlink{0009-0005-8110-397X}
\and Lacroix, L. \inst{4, 8}
\and M\"uller-Bravo, T. E. \inst{12, 13} \orcidlink{0000-0003-3939-7167}
\and Nugent, P. \inst{15, 16} \orcidlink{0000-0002-3389-0586}
\and Popovic, B. \inst{1} \orcidlink{0000-0002-8012-6978}
\and Racine, B. \inst{9} \orcidlink{0000-0001-8861-3052}
\and Rosnet, P. \inst{7} \orcidlink{0000-0002-6099-7565}  
\and Rosselli, D. \inst{9} \orcidlink{0000-0001-6839-1421}
\and Sollerman, J. \inst{17} \orcidlink{0000-0003-1546-6615}
\and Terwel, J.H. \inst{3, 18} \orcidlink{0000-0001-9834-3439}
\and Townsend, A. \inst{5} \orcidlink{0000-0001-6343-3362}
\and Brugger, J. \inst{19} 
\and Bellm, E. C. \inst{20} \orcidlink{0000-0001-8018-5348}
\and Kasliwal, M. M. \inst{21} \orcidlink{0000-0002-5619-4938}
\and Kulkarni, S. \inst{21}
\and Laher, R. R. \inst{22} \orcidlink{0000-0003-2451-5482}
\and Masci, F. J. \inst{22} \orcidlink{0000-0002-8532-9395}
\and Riddle, R. L. \inst{19}
\and Sharma, Y. \inst{21} \orcidlink{0000-0003-4531-1745}
          }

    \institute{Univ Lyon, Univ Claude Bernard Lyon 1, CNRS, IP2I Lyon/IN2P3, UMR 5822, F-69622, Villeurbanne, France
         \and Department of Physics, Lancaster University, Lancs LA1 4YB, UK
         \and School of Physics, Trinity College Dublin, College Green, Dublin 2, Ireland
          \and Oskar Klein Centre, Department of Physics, Stockholm University, SE-10691 Stockholm, Sweden
          \and Institut für Physik, Humboldt Universität zu Berlin, Newtonstr 15, 12101 Berlin, Germany
         \and National Research Council of Canada, Herzberg Astronomy \& Astrophysics Research Centre, 5071 West Saanich Road, Victoria, BC V9E 2E7, Canada
         \and Université Clermont Auvergne, CNRS/IN2P3, LPCA, F-63000 Clermont-Ferrand, France
         \and Sorbonne Université, CNRS/IN2P3, LPNHE, F-75005, Paris, France
         \and Aix Marseille Université, CNRS/IN2P3, CPPM, Marseille, France
         \and Department of Physics, Duke University, Durham, NC 27708, USA
         \and Institute of Astronomy and Kavli Institute for Cosmology, University of Cambridge, Madingley Road, Cambridge CB3 0HA, UK
         \and Institute of Space Sciences (ICE, CSIC), Campus UAB, Carrer de Can Magrans, s/n, E-08193, Barcelona, Spain
         \and Institut d'Estudis Espacials de Catalunya (IEEC), E-08034 Barcelona, Spain
         \and Deutsches Elektronen Synchrotron DESY, Platanenallee 6, 15738, Zeuthen, Germany
         \and Lawrence Berkeley National Laboratory, 1 Cyclotron Road MS 50B-4206, Berkeley, CA, 94720, USA
        \and Department of Astronomy, University of California, Berkeley, 501 Campbell Hall, Berkeley, CA 94720, USA
         \and Oskar Klein Centre, Department of Astronomy, Stockholm University, SE-10691 Stockholm, Sweden
         \and Nordic Optical Telescope, Rambla José Ana Fernández Pérez 7, ES-38711 Breña Baja, Spain
\and Caltech Optical Observatories, California Institute of Technology, Pasadena, CA 91125, USA
\and DIRAC Institute, Department of Astronomy, University of Washington, 3910 15th Avenue NE, Seattle, WA 98195, USA
\and Division of Physics, Mathematics, and Astronomy, California Institute of Technology, Pasadena, CA 91125, USA
\and IPAC, California Institute of Technology, 1200 E. California Blvd, Pasadena, CA 91125, USA
             }

   \date{Received; accepted}
 
  \abstract
   {Type Ia supernova (SN Ia) cosmology will soon be dominated by systematic, rather than statistical, uncertainties, making it crucial to understand the remaining unknown phenomenon that might affect their luminosity, such as astrophysical biases. To be used in cosmology, SN Ia magnitudes need to be standardised, i.e. corrected for their correlation with lightcurve width and colour.} 
   {Here we investigate how the standardisation procedure used to reduce the scatter of SN Ia luminosities is affected by their environment, with the aim to reduce scatter and improve standardisation.} 
   {We first study the SN Ia stretch distribution, as well as its dependence on environment, as characterised by local and global $(g-z)$ colour and stellar mass. We then look at the standardisation parameter $\alpha$, which accounts for the correlation between residuals and stretch, along with its environment dependence and linearity. We finally compute magnitude offsets between SNe in different astrophysical environments after colour and stretch standardisation, aka steps. This analysis is made possible due to the unprecedented statistics of the volume-limited Zwicky Transient Facility (ZTF) SN Ia DR2 sample.}
   {The stretch distribution exhibits a bimodal behaviour, as previously found in literature. However, we find the distribution to be dependent on environment. Namely, the mean stretch modes decrease with host stellar mass, at a $9.2\sigma$ significance. 
   We demonstrate, at the $13.4\sigma$ level, that the stretch-magnitude relation is non-linear, challenging the usual linear stretch-residuals relation currently used in cosmological analyses. Fitting for a broken-$\alpha$ model, we indeed find two different slopes between stretch regimes ($x_1\lessgtr x_1^0$, with $x_1^0=-0.48\pm0.08$): $\alpha_\mathrm{low}=0.271 \pm 0.011$ and $\alpha_\mathrm{high}=0.083 \pm 0.009$, a  $\Delta\alpha=-0.188\pm0.014$ difference. As the relative proportion of SNe Ia in the high-stretch/low-stretch modes evolves with redshift and environment, this implies that a single-fitted $\alpha$ also evolves with redshift and environment. Concerning the environmental magnitude offset $\gamma$, we find it to be greater than $0.12$ mag regardless of the considered environmental tracer used (local or global colour and stellar mass), all measured at the $\geq5\sigma$ level. When accounting for the stretch-non linearity, these steps increase to $\sim0.17$ mag, measured with a 0.01 mag precision. Such strong results highlight the importance of using a large volume limited dataset to probe the underlying SN Ia-host correlations.}
   {}

   \keywords{Cosmology: dark energy -- supernovae: general}

    \titlerunning{ZTF SN Ia DR2: Environmental dependencies of stretch and luminosity}
    \authorrunning{Ginolin, M.}
    \maketitle
%
\section{Introduction}
\label{sec:intro}

Type Ia supernovae (SNe Ia) are standardisable candles that enabled the discovery of the acceleration of the Universe's expansion in the late 1990s 
\citep{Riess_1998, Perlumtter_1999}. Today, they remain a key cosmological probe, as they can uniquely measure the recent ($z<0.5$) expansion rate of the Universe and, as such, are central for the derivation of the dark energy equation of state parameter $w$ \citep{Planck_2020, Pantheonplus}, its potential evolution with redshift, and the direct measurement of the Hubble-Lemaître constant $H_0$ \citep{Freedman_2021,Riess_2022}.

The state-of-the-art measurements of cosmological parameters show that a cosmological constant $\Lambda$ explains the observed properties of dark energy with $w$ compatible with $-1$ at the 3\% precision level \citep{Betoule_2014, Scolnic_2018, Pantheonplus}. However, the direct measurement of $H_0$ is incompatible at the $5\sigma$ level with the standard  model ($\Lambda$ Cold Dark Matter, $\Lambda$CDM) when the parameters are anchored by early Universe physics \citep[][]{Macaulay_2019, Feeney_2019, Riess_2022}. If the latter is not caused by (necessarily multiple, see \citealt{Riess_2022}) sources of systematic uncertainties, this tension would be a sign of new fundamental physics. Yet, no simple theoretical deviation to the fiducial model is able to explain this tension without creating other issues \citep[see][for a recent review]{Schoneberg_2022}. In that context, it is necessary to further investigate the existence of systematic biases that may affect distances derived from SN~Ia data.

SNe~Ia, as "standardisable" candles, have a natural scatter of $\sim0.40$ mag. 
However, two empirical relations, the so-called slower-brighter and 
bluer-brighter relations \citep{Phillips_1993, Tripp_1998}, make use of SN Ia lightcurve properties to reduce that scatter down to $\sim0.15$ mag. The \texttt{SALT} \citep{Guy_2007, Guy_2010, Betoule_2014}  lightcurve fitter is the usual algorithm used to estimate SN~Ia lightcurve stretch $x_1$ and colour parameter $c$ (see \cite{Kenworthy_2021} and Augarde et al. (in prep) for a recent re-coding).

The much reduced scatter provided by these two linear relations makes SNe~Ia the best cosmological distance indicator. Yet, only half of this remaining scatter can be explained by known measurement errors or modelling uncertainties. The rest, dubbed “intrinsic scatter”, may be due to unknown systematic uncertainties that could bias distances, thus the measurement of cosmological parameters.

It has been demonstrated that SN~Ia properties do vary as a function of their environment. The lightcurve stretch, a purely intrinsic property, has been shown to depend on the host environment, such that older and redder environments host on average faster evolving SNe~Ia \citep[e.g.][]{filippenko1989,hamuy1996,Sullivan_2010, Rigault_2020}. Since the galactic star formation quickly evolves with redshift \citep{madau2014}, it has been suggested that SN~Ia intrinsic properties evolve with redshift \citep{howell2007}, as recently demonstrated at the $5\sigma$ level by \cite{Nicolas_2021}. Yet, as long as the stretch brightness dependency is fully captured by the standardisation procedure, such an intrinsic redshift evolution of SN~Ia properties should not affect SN~Ia cosmology, apart for when the actual stretch distribution is needed, e.g. for selection effect bias corrections \citep[][]{scolnic2016}.

However, it has been shown that SNe~Ia from massive hosts are on average brighter after standardisation than these from low-mass hosts \citep[][]{Kelly_2010, Sullivan_2010, Lampeitl_2010, Childress_2013, Rigault_2020}. This so called “mass step” is now accounted for in cosmological analyses \citep{Betoule_2014, Scolnic_2018, Pantheonplus, popovic2024}, but its origin is highly debated. Research works suggest that it could be due to progenitor age \citep[e.g.][]{Rigault_2020, Briday_2021, Kim_2018} or dust property variations \citep[e.g.][]{Brout_Scolnic_2021, Popovic_2021, popovic2023}. Understanding the origin of such variations is required for accurate cosmology, since corrections for redshift evolution or sample selection functions may vary with environment.

In this analysis, we investigate the stretch standardisation procedure, its connection with the mass step, and the relation between SN stretch and progenitor age. In a companion paper \citep{Ginolin_2024b}, we focus on SN~Ia colour, its potential connection with dust, and the accuracy of the colour standardisation. Both papers are based on the second data release of the Zwicky Transient Facility \citep[ZTF, ][]{Bellm_2019,Graham_2019, ZTF, Masci_2019} Cosmology Science Working Group (ZTF SN Ia DR2, \citealt{DR2_overview, Smith_2024}, following the DR1, \citealp{ZTF_DR1}).

This paper starts in Section~\ref{sec:data} with a summary of the ZTF SN Ia DR2 release, where we  introduce the sample selection used to create a well controlled volume-limited dataset. In Section~\ref{sec:stretch_distribution}, we study the stretch distribution, where we present a more complex connection between SN stretch and SN environment than what has been previously reported.
In Section~\ref{sec:stretch_mag_relation}, we investigate in detail the stretch-magnitude relation, that we find to be significantly non-linear. In Section \ref{sec:steps}, we then investigate magnitude offsets due to SN environment, aka steps, as well as their connection to the non-linearity of the stretch-residuals relation. We test the robustness of our results in Section \ref{sec:tests}, discuss our results in Section \ref{sec:discussion}, and conclude with Section~\ref{sec:conclusion}.

Except if mentioned otherwise, we use the recent recalibration of the \texttt{SALT2.4} lightcurve fitter \citep{Guy_2010,Betoule_2014} from \cite{Taylor_2021} as provided by the ZTF SN Ia DR2 release, following \citealt{DR2_overview, Smith_2024}.

\section{Data}
\label{sec:data}

\subsection{Zwicky Transient Facility Cosmology DR2}
\label{sec:ztfdr2}

For this analysis, we use the volume-limited ZTF SN Ia DR2 sample presented in \citealt{DR2_overview, Smith_2024}. 

The initial DR2 sample contains 2663 spectroscopically confirmed SNe~Ia passing basic quality cuts:
(1) Good lightcurve sampling, i.e., at least seven $5\sigma$ flux detections within the $-10$ to $+40$ days rest-frame phase range, with at least 2 pre-max detections, at least 2 post-max detections, and at least detections in two bands, (2) Stretch $x_1 \in [3, 3]$ measured with a precision better than $\sigma_{x_1}=1$, (3) Colour $c\in[-0.2, 0.8]$ measured with a precision better than $\sigma_c=0.1$, (4) Precision on the estimated peak-luminosity time better than $\sigma_{t_0}=1$ day and (5) \texttt{SALT2} lightcurve fit probability greater than $10^{-7}$.
Finally, to have a volume limited sample, we limit ourselves to SNe Ia having a redshift $z<0.06$, following the prescription from survey simulations \citep{Amenouche_2024}, so that our sample is free from significant non-random selection functions. SNe Ia in the volume-limited sample thus probe the underlying SN~Ia population, with no need to model for complex selection function bias correction.

As suggested by \cite{Rose_2022}, we extend the colour range further than the usual literature cut, from $c<0.3$ to $c<0.8$, as we have a significant fraction (10\%) of red SNe~Ia. In agreement with \cite{Rose_2022}, we see no behaviour evolution of SNe at $c>0.3$, and we thus apply a less restrictive cut (see detailed study in \citealt{Ginolin_2024b}). In Sect.~\ref{sec:tests}, we show that only considering objects with $c<0.3$ has no significant impact on our results.

We also discard SNe~Ia classified by the ZTF SN Ia collaboration as peculiar. As detailed in \cite{Dimitriadis_2024}, SNe Ia typically classified as peculiar are the 91bg or Ia-CSM subclasses. We however keep the SNe Ia 91t, as they usually pass cosmological cuts. We show in Sect.~\ref{sec:tests} that including the 91bg or discarding the 91t from our sample has no significant impact on our results.
We further discard $26$ objects with missing host photometry (see Sect. \ref{sec:environmental_properties}). 

The final volume-limited sample is thus comprised of 945 SNe~Ia. Of these, 75\% have a redshift coming from host spectral features, mostly from the MOST Hosts Dark Energy Spectroscopic Instrument (DESI) program \citep{MOST_Hosts}, with a typical precision of $\sigma_z\leq10^{-4}$, while 25\% have a redshift derived from SN Ia spectral features. As demonstrated in \cite{Smith_2024}, these SN Ia-features redshifts are unbiased and have a typical precision of $\sigma_z\leq3\times10^{-3}$.

As mentioned in \cite{DR2_overview}, a non-linearity in the ZTF CCD read-out started to affect the data in November 2019, following an update of the CCD waveforms. This effect, dubbed  "pocket effect", is described fully in Lacroix et al. (in prep), and will be corrected for in the upcoming ZTF SN Ia DR2.5. The pocket effect impacts the point spread function, and the amplitude of this effect depends on the signal to noise of a given exposure. The overall effect is of the order of 1\% between 15 mag and 19 mag and is colour independent. This is an issue for cosmology, as it prevent us from deriving accurate absolute fluxes, but does not affect the ZTF SN Ia DR2 analysis, as we only use it for self-comparison. However, simulations have shown that, for this DR2, the pocket effect only marginally affects the stretch, while leaving the colour and the Hubble residuals unchanged. For SNe Ia affected by the pocket effect, the stretch $x_1$ is shifted by $\Delta x_1=-0.1$, half of the typical $x_1$ error in the DR2, independently of the true $x_1$ (cf \citealt{DR2_overview}). 
We show in Sect. \ref{sec:pocket_effect} that our conclusions are not impacted by the pocket effect, since our results do not significantly vary when splitting our sample between SNe Ia acquired pre- and post-November 2019.

\subsection{Local and global host properties}
\label{sec:environmental_properties}

To study the correlation of SN properties with environment, we use the four environmental tracers available in the DR2: stellar mass and colour (\texttt{ps1.g-ps1.z} from PanSTARRS \citealt{Pan-STARRS}), both local (2 kpc radius around the SN) and global (whole host galaxy). Environmental property estimation is done using the \texttt{HostPhot} package \citep{HOSTPHOT}. It is described in further detail in \cite{Smith_2024}, along with the parameter distributions.
When comparing SNe Ia from these environments, we split them into two subsamples, using the following cuts:
\begin{itemize}
    \item Global mass: we take the standard literature cut $\log(M_\star/M_\odot)_\mathrm{cut}^\mathrm{global}=10$.
    \item Local mass: we take the median of the local mass distribution $\log(M_\star/M_\odot)_\mathrm{cut}^\mathrm{local}=8.9$.
    \item Local and global colour: we take the gap visible in the bimodal host colour distribution $(g-z)_\mathrm{cut}=1$ mag.
\end{itemize}

\section{Stretch distribution}
\label{sec:stretch_distribution}

In this section, we study the distribution of the \texttt{SALT2.4} standardisation parameter $x_1$ (stretch). All fits are done through likelihood minimisation, with the use of \texttt{iminuit} \citep{iminuit}. 

\subsection{The nearby SN~Ia stretch distribution}
\label{sec:stretch_1D}

The ZTF SN Ia DR2 sample stretch ($x_1$) distribution is shown in the upper panel of Fig.~\ref{fig:stretch_distrib}. It exhibits a clear bimodal shape, with a low-stretch mode at $x_1\sim-1.2$ and a high-stretch mode at $x_1\sim 0.4$, the first mode being approximately twice more populated than the second. As an additional test, we compute differences in Akaike Information Criterion (AIC, \cite{AIC}) with other distributions that could match the stretch distribution by eye. The double Gaussian is strongly favoured over a single Gaussian ($\Delta\mathrm{AIC}=105$) and a skewed Gaussian ($\Delta\mathrm{AIC}=40$).

\subsubsection{Discussion on SN~Ia stretch bimodality}
\label{sec:stretch_1D_bimodality}

This distribution is very similar to the one from the Nearby Supernova Factory (SNfactory, \citealt{Aldering_2002, Rigault_2020}) dataset studied in detail in \citet[][hereafter \citetalias{Nicolas_2021}]{Nicolas_2021}. It is also similar to other nearby SN Ia data sets, like the one from Foundation \citep{Foley_2018} and the low-$z$ Pantheon compilation~\citep{Scolnic_2018} both studied in Figure 6 of \cite{Popovic_2021}, or the Supercal compilation \citep{Scolnic_2015} studied in Figure 3 of \cite{Wojtak_2023}. These latter samples do exhibit a bimodal distribution, but some with both modes equally populated, unlike what we observed with our volume-limited sample. 
This is likely caused by complex selection effects affecting those low-$z$ samples (see discussion in \citetalias{Nicolas_2021}). 
In contrast, higher redshift samples do not display a clear bimodal distribution. According to \citetalias{Nicolas_2021} and \cite{Rigault_2020}, this is to be expected. In their model, since low-stretch SNe~Ia only originate from old environments, and since the cosmic star formation strongly increases with redshift \citep{madau2014, Tasca_2015}, the fraction of (supposedly) old progenitor SNe~Ia is lower and, consequently, the low-stretch  mode tends to vanish with redshift.

\begin{figure}
    \centering
    \includegraphics[width=1\columnwidth]{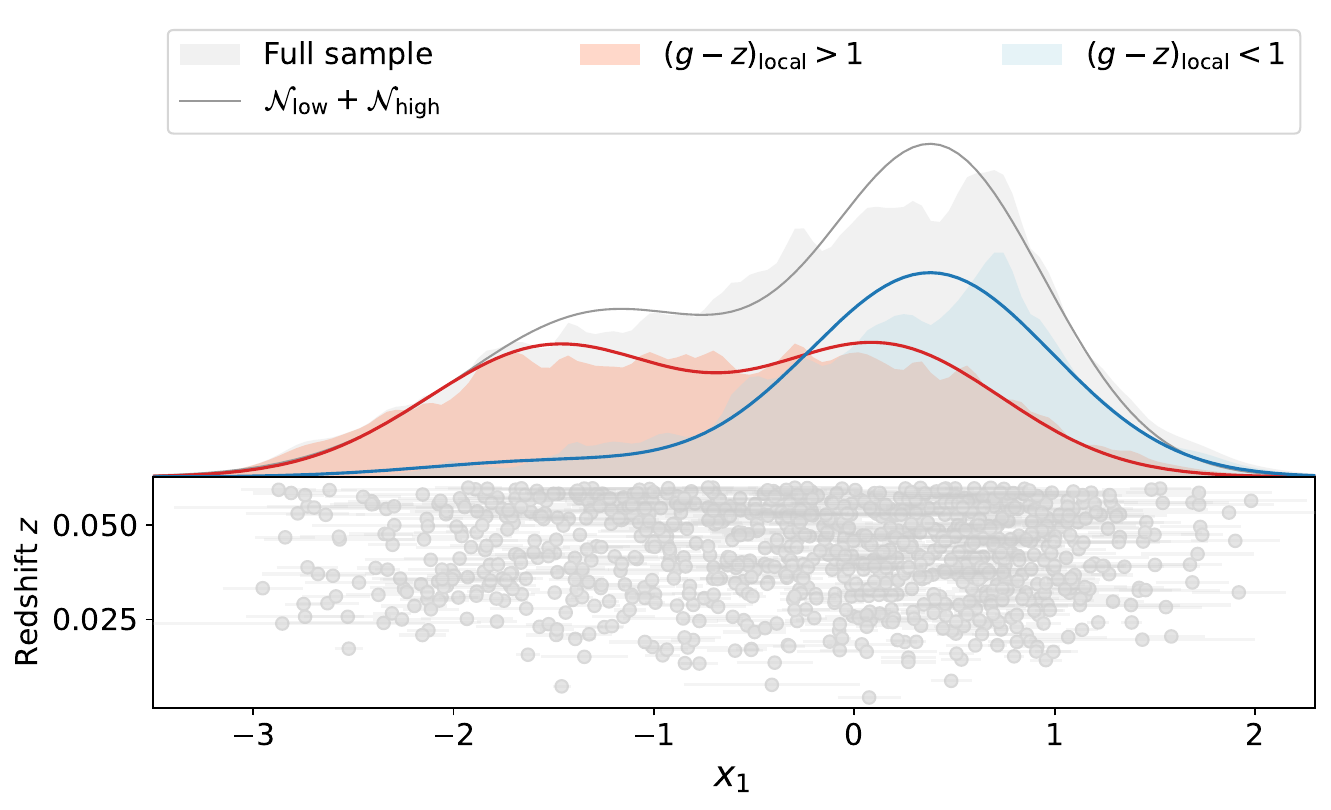}
    \caption{
    \textit{Top:} Ideogram of the stretch distribution for the full sample (in grey), and for SNe in locally red/blue environments (509/429 SNe). The full grey line is the bimodal Gaussian from \citetalias{Nicolas_2021} described in Sect.~\ref{sec:stretch_distribution}, while the red/blue lines are bimodal Gaussian fits to the old/young progenitor subpopulations.
    \textit{Bottom:} Stretch vs redshift $z$. The fact that there is no depletion of low stretch SNe at a higher redshift is a consequence of the volume-limited cut.}
    \label{fig:stretch_distrib}
\end{figure}

We present in Table~\ref{tab:stretch_n21} the bimodal Gaussian distribution parameters estimated on the volume-limited ZTF data set, and the best fit distribution is shown in Fig.~\ref{fig:stretch_distrib}. The fitted model is defined as follows: $P(x_1) = r\mathcal{N}(x_1\,|\,\overline{x_1^\mathrm{high}}, \sigma_\mathrm{high}^2) + (1-r)\mathcal{N}(x_1\,|\,\overline{x_1^\mathrm{low}}, \sigma_\mathrm{low}^2)$.

\begin{table}
\centering
\small
\caption{Best-fit values for the stretch model presented in Sect. \ref{sec:stretch_1D_bimodality}.}
\begin{tabular}{l c c c} 
\hline\\[-0.8em]
\hline\\[-0.5em]
Param. & \citetalias{Nicolas_2021} (fiducial) & This work & Difference ($\sigma$)\\[0.15em]
\hline\\[-0.5em]
 $\overline{x_1^\mathrm{high}}$ & $0.37\pm0.05$ &  $0.42\pm0.08$ & 0.5 \\[0.30em]
$\sigma_\mathrm{high}$ & $0.61\pm0.04$ & $0.54\pm0.05$ & 1.2 \\[0.30em]
$\overline{x_1^\mathrm{low}}$ & $-1.22\pm0.16$ & $-1.24\pm0.18$ & 0.1 \\[0.30em]
$\sigma_\mathrm{low}$ & $0.56\pm0.10$ &$0.73\pm0.09$ & 1.3 \\[0.30em]
$r$ & $0.755\pm0.05$ & $0.59\pm0.07$ & 1.9\\[0.3em]
\hline
\end{tabular}
\label{tab:stretch_n21}
\tablefoot{Means ($\overline{x_1}$) and standard deviations ($\sigma$) of the two Gaussian modes of the bimodal stretch model. $r$ is the relative amplitude of each mode.}
\end{table}

Our results are in good agreements with those reported by \citetalias{Nicolas_2021}. To get the ratio of the two modes $r$ for \citetalias{Nicolas_2021}, we had to use the assumption made in \cite{Rigault_2020} that at their redshift, the fraction of young and old progenitors is half-half. The stretch mode parameters are compatible at the $\sim1\sigma$ level. The sole remarkable difference is that the amplitude of the $r$ parameter seems lower in our data set, though only at a $1.9\sigma$ level. If confirmed, such a reduced amplitude of the high-stretch model could suggest that either the low-stretch mode is slightly more populated than the high-stretch one in the old (delayed) population, or that the fraction of old population is slightly higher than the expected 50\% modelled by \cite{Rigault_2020} for our median redshift ($z_\mathrm{median}=0.044$).

\subsubsection{Stretch distribution in locally red or blue environments}
\label{sec:stretch_1D_env}

We split the ZTF volume-limited SNe Ia as a function of their local environment to further investigate the origin of the observed bimodality. Following e.g. \cite{Roman_2018, Kelsey_2023, Briday_2021, Wiseman_2022}, we use the local (2 kpc-radius) colour $(g-z)_\mathrm{local}$, as it performs well as a proxy for the underlying SN~Ia prompt and delayed subpopulations used in \citetalias{Nicolas_2021}, with locally blue/red environments hosting young/old SN Ia progenitors. Splitting the ZTF data at $(g-z)_\mathrm{local} = 1\,\mathrm{mag}$, 54\% of the SNe Ia are found in a locally red environments and 46\% in a locally blue environment. 

The stretch distribution per local environment and the best bimodal fit for each subgroup is shown in the top panel of Fig.~\ref{fig:stretch_distrib}. We draw three conclusions from this figure.

First, the locally red environment SNe Ia are consistent with being equally populated with each mode ($r=49\pm9\%$). 
Second,  locally blue environment SNe~Ia have a non-null low-stretch mode ($7.5\pm1.8\%$).
This is consistent with \citetalias{Nicolas_2021} modelling, that assumes the low-stretch mode to only be accessible to old-population SNe~Ia once the environmental contamination is accounted for. Indeed, according to \cite{Briday_2021}, $13^{+7}_{-6}\%$  of the colour classifications are false, i.e. old-progenitor SNe Ia are associated to blue environments and vice versa. We thus expect 8\% of old-population SNe~Ia to be classified as locally blue (see Fig.~3 of \cite{Briday_2021}), and so $\sim4\%$ of the locally blue sample to be in the low-stretch mode.
Third, the mean of the high-stretch mode seems to slightly vary as a function of local environment, with $\Delta \overline{x_1^\mathrm{high}}=0.24\pm0.15\,(1.6\sigma)$. This is further discussed in the next section.

\subsection{Correlation between SN stretch and SN environment}
\label{sec:stretch_2D}

In this subsection, we investigate the correlation of stretch with environment, namely local colour $(g-z)$ and global mass. We are able to disentangle their effects, with the proportion of SNe Ia in high/low stretch modes being dependent on local colour, while the dependence of the mean stretch is tied to global mass. This is then modelled in Sect. \ref{sec:stretch_model}.

\subsubsection{Qualitative study of stretch environmental correlations}
\label{sec:stretch_2D_quantitative}

\begin{figure}
   \centering
   \includegraphics[width=0.9\columnwidth]{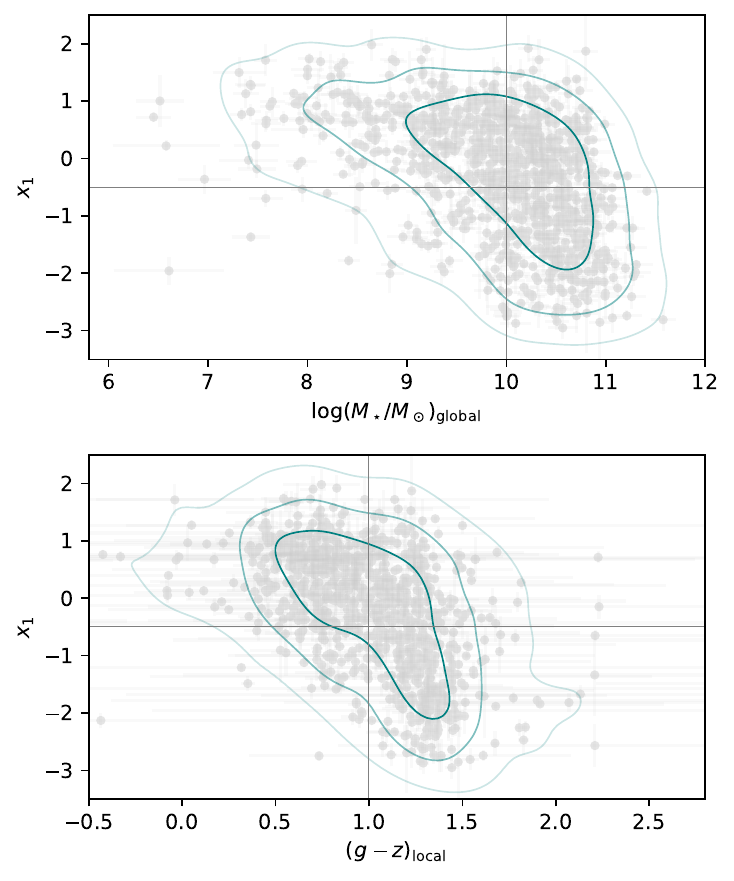}
    \caption{Correlation between SN~Ia lightcurve stretch ($x_1$) and host stellar mass (\emph{top}), local environmental colour (\emph{bottom}). The contours show the area containing 97\%, 84\%, and 50\% of the SNe Ia. Vertical lines show the environment splitting values, while the horizontal lines show $x_1=-0.5$, the typical transition point between stretch modes (see Fig.~\ref{fig:stretch_distrib}). 
    }
        \label{fig:stretch_mass_lcolour}
\end{figure}

\begin{figure*}
   \centering
   \includegraphics[width=2\columnwidth]{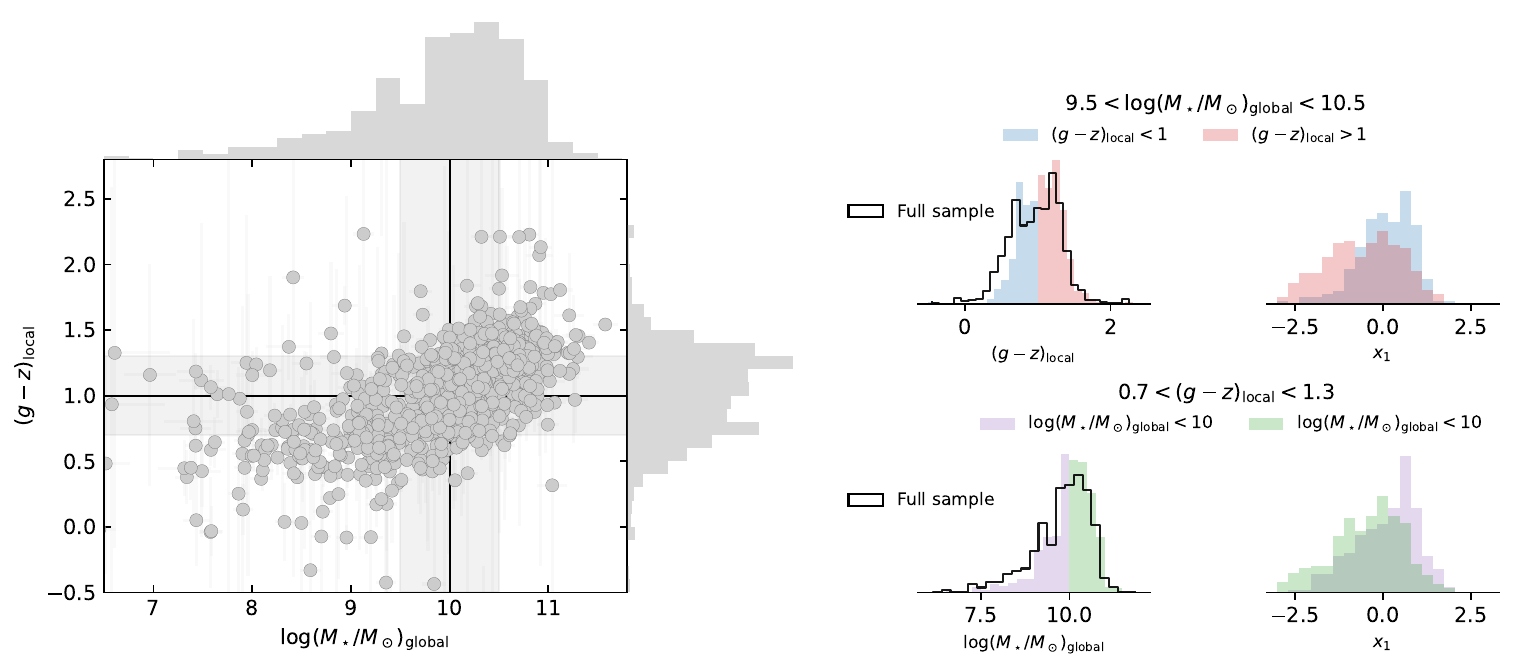}
      \caption{Connections between SN~Ia lightcurve stretch ($x_1$), global stellar mass ($\log(M_*/M_\odot)_\mathrm{global}$) and local environmental colour ($(g-z)_\mathrm{local}$), illustrating the complex correlation between stretch and SN environmental properties. 
      \emph{Left}: Local colour vs. global host stellar mass. The light grey band shows the 25\%-75\% percentile range for host global mass (vertical, [9.5, 10.5] dex) and local environmental colour (horizontal, [0.7, 1.3] mag). These cuts are used to select SNe~Ia shown in the right panels. 
      \emph{Top-right:} Local environmental colour and SN stretch distributions for SNe Ia whose host global mass is in [9.5, 10.5] dex. Blue and red histograms show locally blue and red environment SNe Ia. The black histogram is the local colour distribution for the full sample. 
      \emph{Bottom-right:} Global host stellar mass and SN stretch distributions for SNe~Ia whose local environmental colour is in [0.7, 1.3] mag. Purple and green histograms show low and high mass hosts SNe Ia.
      }
      \label{fig:stretch_env_origin}
\end{figure*}

We show in Fig.~\ref{fig:stretch_mass_lcolour} the connection between SN~Ia lightcurve stretch parameter $x_1$ and host galaxy mass as well as local environmental colour. We clearly see that the low-stretch mode only exists in locally red environments and massive host galaxies. Both are connected, since these environmental parameters are highly correlated, as illustrated in Fig.~\ref{fig:stretch_env_origin}. This is consistent with earlier findings that the low-stretch mode only exists in old stellar population progenitors \citep[e.g.][]{hamuy1996,howell2007,Rigault_2020,Nicolas_2021, Larison_2024}, as those strongly favour massive hosts and make the local environment redder.

Figure~\ref{fig:stretch_mass_lcolour} further shows that the connection between stretch and environment is more complex than the simple appearance of a low stretch mode in old-population environments as modelled, for instance, by \cite{Nicolas_2021}. There seems to be a correlation, clearly visible in the high-stretch mode, such that lower mass host/bluer environments tends to have a higher stretch. 

To investigate the origin of this effect, we look at the detailed connection between SN lightcurve stretch, global host mass and local environmental colour in Fig.~\ref{fig:stretch_env_origin}. 

We split the data along host stellar mass and local environment colour, to investigate which environmental parameters is the most connected the our observations, namely the appearance of the low-stretch mode and the stretch evolution. To do that, we only consider SNe~Ia within the 25\%-75\% range of an environmental parameter (for example host stellar mass), and then study the correlation between $x_1$ and the other environmental tracer (e.g. local colour), as shown on the top-right panels of Fig.~\ref{fig:stretch_env_origin}. We see that the local colour distribution post host stellar mass 25\%-75\% selection is similar to that of the entire initial data set. Looking at the bottom-right panel of Fig.~\ref{fig:stretch_env_origin}, we notice the same thing for the stellar mass distribution after the local-environmental 25\%-75\% selection. However, the resulting $x_1$ distributions per environment are very informative on the origin of the two observed effect. For the given stellar mass range, locally red SNe~Ia seem to have the same high-stretch mode and show in addition the apparition of the low-stretch mode (top-right panel). For a given local environmental colour range, high and low mass host SNe~Ia have a similar but shifted ($\Delta x_1\sim-0.5$) $x_1$ distributions (bottom-right panel). 

From these observations, we conclude that the proportion of SNe in each stretch modes is directly connected to the local colour, while the apparent evolution of the mean stretch is connected to the global host stellar mass. The observed evolution of the high-stretch mode in the bottom panel of Fig.~\ref{fig:stretch_mass_lcolour} thus seems to be a projection of the evolution connected to the global host stellar mass. 
Then, assuming that the lightcurve stretch is an intrinsic SN~Ia property, we interpret these observations as follows: the low-stretch mode is only accessible to old-population progenitors (delayed) as already suggested in the literature \citep{howell2007,Sullivan_2010,Rigault_2020, Nicolas_2021}, but the stretch evolution is likely connected to the progenitor metallicity, since stellar mass and stellar metallicity are tightly correlated \citep{Tremonti_2004, Sanchez_2017}. This may be connected to literature observations that SN~Ia stretch is directly linked to the progenitor mass or produced $^{56}$Ni mass \citep[e.g.][and references therein]{Dhawan_2017, scalzo2014}.

The link with SN stretch and environment is also studied with the ZTF SN Ia sample using clusters \citep{Ruppin_2024} and voids \citep{Aubert_2024} as environment tracers.

\subsubsection{Origin of the shifted stretch distribution between host redshift and SNID redshift SNe~Ia}

\begin{figure}
   \centering
   \includegraphics[width=1\columnwidth]{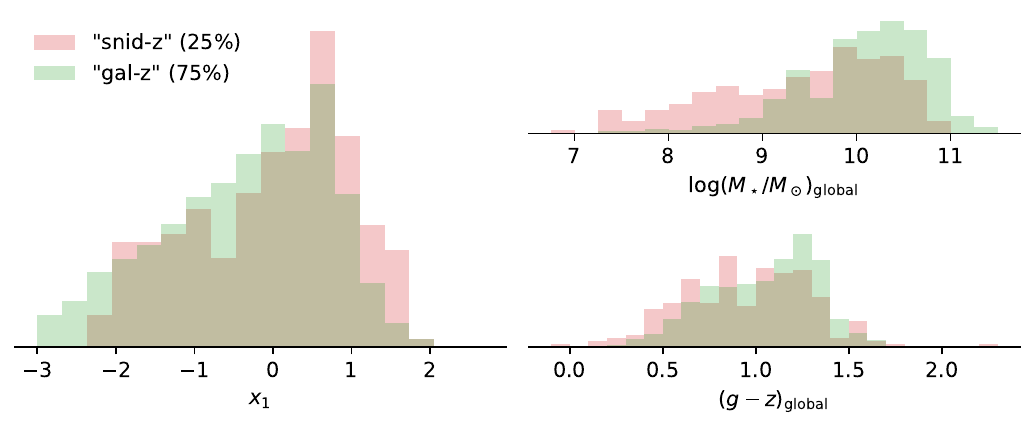}
    \caption{SN~Ia lightcurve stretch ($x_1$), global host mass ($\log(M_*/M_\odot)_\mathrm{global}$) and local environmental colour ($(g-z)_\mathrm{local}$) distributions split per redshift origin. SNe~Ia with galaxy redshifts are plotted in green ($\sigma z \sim 10^{-4}$), and SNe Ia relying on low-resolution SN spectra (extracted from SEDm, \cite{SEDm}), either from SNID \citep{SNID} or on host emission lines are plotted in red ($\sigma z \sim 10^{-3}$). The apparent stretch offset is explained by the selection function associated with galaxy redshifts, which favours high-mass hosts.}
        \label{fig:stretch_zsource}
\end{figure}

We present in Fig.~\ref{fig:stretch_zsource} the SN~Ia stretch distribution comparing SNe Ia having a host galaxy redshift (“gal-z”) with those without, where the redhsift is extracted from the low resolution SN spectra from the Spectral Energy Distribution machine (SEDm,  \citealt{SEDm}), either from SNID \citep{SNID} or from host emission lines ("snid-z”, see \citealt{Smith_2024} for details). 

We note that the stretch distribution for “gal-z" SNe~Ia is shifted by $\Delta x_1\sim0.5$ in comparison to that from "snid-z" SNe~Ia. This shift is explained by the selection function caused by our galaxy redshift sources (e.g. the (extended) Baryon Oscillation Spectroscopic Survey ((e)BOSS), see \citealt{Smith_2024}), which strongly favours massive hosts, as illustrated in the top-right panel of Fig.~\ref{fig:stretch_zsource}. Interestingly, the local colour distributions are similar between these two SN samples. This further supports our claim that the stretch shift as a function of environment is driven by a physical mechanism more directly connected to global stellar mass than local colour, like stellar metallicity (see details in Sect.~\ref{sec:stretch_2D_quantitative}).

\subsubsection{Modelling of the stretch environmental dependencies}
\label{sec:stretch_model}

To quantify the observations described in Sect.~\ref{sec:stretch_2D_quantitative}, we extend the stretch model from \citetalias{Nicolas_2021} to account for the global mass dependency of the high/low stretch modes mean $\overline{x_1^\mathrm{high/low}}$ described in the previous section, and to get a more robust modelling of the relative fraction of high-stretch SNe~Ia $r$ as a function of the local colour. We denote $(g-z)_\mathrm{local} \equiv c_\mathrm{env}$ and $\log(M_\star/M_\odot)_\mathrm{global} \equiv M_\mathrm{env}$ for readability. The model is the following:
\begin{align}
    \label{eq:stretch2D_1}
    P\left(x_1\,|\,c_\mathrm{env}, M_\mathrm{env}\right)=r(c_\mathrm{env})&\mathcal{N}(x_1|\overline{x_1^\mathrm{high}}(M_\mathrm{env}), \sigma_\mathrm{high}^2) + \\ \nonumber
    \left(1-r(c_\mathrm{env})\right)&\mathcal{N}(x_1|\overline{x_1^\mathrm{low}}(M_\mathrm{env}), \sigma_\mathrm{low}^2) 
\end{align}
with:
\begin{align}
    \label{eq:stretch2D_2}
    r(c_\mathrm{env})&=r_\mathrm{red} + (r_\mathrm{blue}-r_\mathrm{red})\times \mathcal{S}\left(\frac{1}{K_c}[c_\mathrm{env}-c_\mathrm{env}^0]\right)\\ 
    \label{eq:stretch2D_3}
    \overline{x_1^\mathrm{high/low}}(M_\mathrm{env})&=K_M\times (M_\mathrm{env}-10)+\overline{x_1^\mathrm{high/low, 0}}
\end{align}

\begin{figure}
   \centering
   \includegraphics[width=0.9\columnwidth]{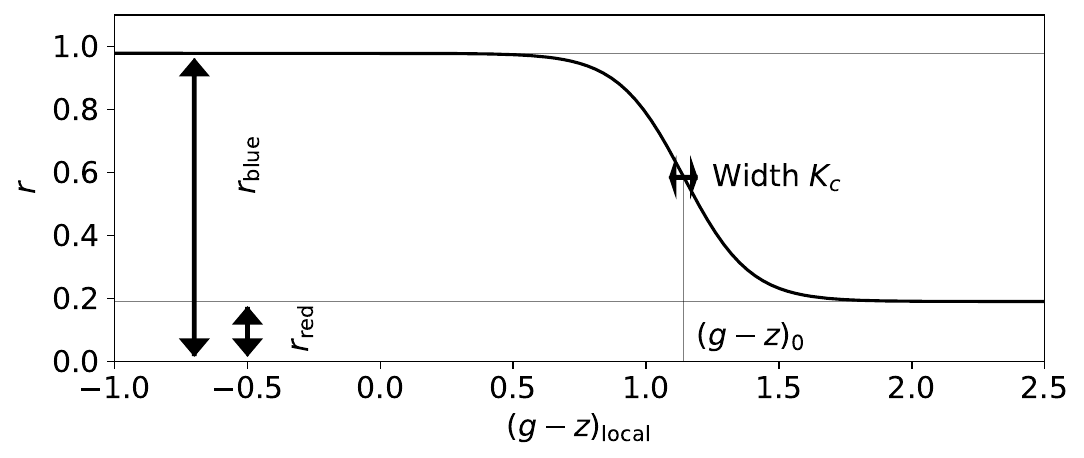}
      \caption{Fraction of SNe in the high stretch mode as a function of the local colour, as modelled in Eq. \ref{eq:stretch2D_2}.}
         \label{fig:stretch_phigh}
\end{figure}

In Equation~\ref{eq:stretch2D_2}, $\mathcal{S}$ is a sigmoid function, and $r_\mathrm{red}$ and $r_\mathrm{blue}$ correspond to the fraction of high-stretch mode SNe~Ia in the blue and red ends of the environmental colour distribution, as illustrated in Fig. \ref{fig:stretch_phigh}. 

We fit the data with this model accounting for errors on $x_1$ and $c_\mathrm{env}$ (i.e., the local colour $(g-z)_\mathrm{local}$). The best fit parameters are displayed in Table~\ref{tab:stretch_2D}. 
To quantify the improvement of this model over simpler versions, we compute AIC differences. Comparing \citetalias{Nicolas_2021}'s model, where the transition between the young and old progenitors stretch distribution (blue and red distribution in Fig. \ref{fig:stretch_distrib}) is sharp, with a model with a smooth transition (the sigmoid function plotted in Fig. \ref{fig:stretch_phigh}), we find $\Delta\mathrm{AIC}=29$, strongly favouring the smooth transition. We then compare this model to the full model, in which the means of the stretch modes evolves with mass, we get $\Delta\mathrm{AIC}=76$. The addition of both a continuous dependence of the fraction of SNe Ia in the high-stretch mode on local colour and of the evolution of the means of the stretch modes with global mass is thus justified, as it is strongly supported by the data.

The linear global mass dependency of the stretch modes ($K_M$) is non-zero at the $9.2\sigma$ level, with $K_M=-0.30 \pm 0.03$ dex$^{-1}$, strongly supporting the evidence that the means of the stretch modes are environment dependent.

\begin{table}
\centering
\small
\caption{Best fitted parameters for $P(x_1 | c_\mathrm{env}, M_\mathrm{env})$ from Eq. \ref{eq:stretch2D_1}-\ref{eq:stretch2D_3}.}
\begin{tabular}{r c} 
\hline\\[-0.8em]
\hline\\[-0.5em]
Parameter & Value\\[0.15em]
\hline\\[-0.5em]
$K_M$ & $-0.30\pm0.03$\\[0.30em]
$\overline{x_1^\mathrm{high, 0}}$ & $0.23\pm0.04$ \\[0.30em]
$\sigma_\mathrm{high}$ & $0.54\pm0.03$\\[0.30em]
$\overline{x_1^\mathrm{low, 0}}$ & $-1.34\pm0.07$ \\[0.30em]
$\sigma_\mathrm{low}$ & $0.64\pm0.05$\\[0.30em]
$r_\mathrm{red}$ & $0.19\pm0.07$\\[0.30em]
$r_\mathrm{blue}$ & $0.98\pm0.03$\\[0.30em]
$(g-z)_\mathrm{local}^0$ & $1.14\pm0.04$\\[0.30em]
$K_c$ & $-0.124\pm0.028$\\[0.30em]
\hline
\end{tabular}
\label{tab:stretch_2D}
\end{table}

We also note that the apparition of the low-stretch mode is progressive, and the fraction of SNe Ia in the low-stretch mode only reaches its redder value of $(1-r)\sim0.8$ at $(g-z)_\mathrm{local}\sim1.7$ mag, as visible in Figure \ref{fig:stretch_phigh}. This behaviour is more complex than the sharp transition modelled in \citetalias{Nicolas_2021}, and explains the $\sim50\%$ of low-stretch SNe Ia in red environments seen in Fig. \ref{fig:stretch_distrib}.

\section{Stretch-residuals relation}
\label{sec:stretch_mag_relation}

In this section, we study the SN~Ia stretch standardisation procedure accounting for the brighter-slower \cite{Phillips_1993} relation. We briefly introduce SN~Ia standardisation in Sect.~\ref{sec:standardisation}, and present our fitting procedure in Sect. \ref{sec:totalchi2}, as well as the resulting standardisation parameters in Sect. \ref{sec:result_standardisation}. We then assess the universality of this procedure as a function of SN environment in Sect.~\ref{sec:alpha_env}, to finally question its assumed linearity in Sect.~\ref{sec:alpha_linearity}.

The colour standardisation is studied in detail in a companion paper \citep{Ginolin_2024b}.

\subsection{SNe Ia standardisation}
\label{sec:standardisation}

We define the difference between observed and modelled distance moduli, also called Hubble residuals, as:
\begin{equation}
      \Delta\mu = \mu_{\text{obs}}-\mu_{\text{cosmo}},
 \end{equation}
  where $\mu_{\text{cosmo}}$ is calculated in \texttt{astropy} \citep{astropy:2013, astropy:2018}, following a flat $\Lambda$CDM cosmology given by \citet[][$\Omega_m=0.315$]{Planck_2020}, plus a blinded magnitude offset. We then use Chauvenet's criterion to reject outliers of the cosmological fit, which discards 7 SNe.

The usual standardisation formula for SNe Ia is given by:
\begin{align}
\label{eq:standardisation}
      \mu_{\text{obs}} &= m_B-M_0 - \beta c + \alpha x_1 -\gamma p + \Delta_b \\
      &= \mu - (\beta c -\alpha x_1 +\gamma p -\Delta_b) \nonumber
\end{align}
 where $M_0$ is the absolute B-band SN magnitude (degenerate with $H_0$), $\alpha$ and $\beta$ are the linear standardisation coefficients correcting for the stretch ($x_1$) and colour ($c$) SN~Ia variations following the slower-brighter and bluer-brighter relations \citep{Tripp_1998}. 
 The $\gamma p$ term accounts for SN~Ia magnitude environmental dependencies \cite[e.g.][]{Kelly_2010, Sullivan_2010, Briday_2021}. $p$ is the probability that an environmental tracer $m$ is below a given splitting value ("cut", see Table.~\ref{tab:alpham0_env}), while  $\gamma$ is the magnitude offset between the SN~Ia subpopulations below or above the environment cut, aka the "step".
 In practice, $p$ ($\in[0,1]$) is the cumulative distribution function of the environmental proxy $m$ measured with an error $\delta m$ evaluated at the cut value, such that $p=\int_{-\infty}^\mathrm{cut}\mathcal{N}(x,m, \delta m)dx$.
 In recent cosmological analyses, $m$ is usually the global host stellar mass and $\gamma$ is referred to as the mass step.
 
The $\Delta_b$ term accounts for selection function affecting the survey, since overly bright objects are easier to acquire and to classify. The correct modelling procedure of this term is highly discussed, and has been shown to bias the environmental correction if not accurately done \cite[e.g. ][]{Smith_2020, Popovic_2021,Nicolas_2021,Wiseman_2022}. In this analysis, to avoid such complications, we use the volume-limited sample of the ZTF DR2 sample ($z<0.06$), which is free from non-random selection functions either from lightcurve estimation or spectral typing \citep{Smith_2024, Amenouche_2024}. Consequently, we set $\Delta_b=0$.
 
The standardisation (i.e. estimation of $M_0$, $\alpha$, $\beta$, $\gamma$, cf. Eq. ~\ref{eq:standardisation}) is done using total-$\chi^2$ minimisation. 
The total-$\chi^2$ approach enables to  fit a model explaining an observed $y$-variable (here $\Delta\mu$) that depends on input noisy $x$-variables (here $c$, and $x_1$, $p$). To do so, and unlike simple $\chi^2$ minimisation, one has to fit for the true (noisy) $x$-variables that are used, in turn, to estimate $y$.

\subsection{Total-$\chi^2$ minimisation.}
\label{sec:totalchi2}

In practice, for a sample containing N SNe, we fit for $3\times N + 5$ parameters: the $3\times N$ parameters corresponding to the true values of the observed noisy standardisation variables ($c^\mathrm{true}$, $x_1^\mathrm{true}$ and $p^\mathrm{true}$), the four standardisation parameters of interest from Eq. \ref{eq:standardisation} ($\alpha$, $\beta$, $\gamma$ and $M_0$), plus an intrinsic magnitude scatter $\sigma_\mathrm{int}$ to account for leftover dispersion in the residuals.

The “total” $\chi^2$ is thus the sum of two parts, one quantifying the likelihood that the observed $x$ correspond to the fitted $x_\mathrm{true}$:
\begin{equation}
    \chi^2_\mathrm{param}=\sum_{i}\frac{(x_1^\mathrm{true,i}-x_1^i)^2}{{\sigma_\mathrm{x_1}^i}^2} + \sum_{i}\frac{(c^\mathrm{true,i}-c^i)^2}{{\sigma_\mathrm{c}^i}^2} + \sum_{i}\frac{(p^\mathrm{true,i}-p^i)^2}{{\sigma_\mathrm{p}^i}^2}
\end{equation}

and the usual $\chi^2$ standardisation, computed with the "true" $x$-variables:

\begin{equation}
    \chi^2_\mathrm{res}=\sum_{i}\frac{(\mu^i-\beta c^\mathrm{true, i}+\alpha x_1^\mathrm{true,i}+\gamma p^\mathrm{true,i})^2}{{\sigma_\mathrm{obs}^i}^2+\sigma_\mathrm{int}^2} 
\end{equation}

where ${\sigma_\mathrm{obs}^i}^2=\begin{pmatrix} 1 & \alpha & \beta \end{pmatrix} \mathbb{C}^i \begin{pmatrix} 1 \\ \alpha \\ \beta \end{pmatrix}$, and $\mathbb{C}^i$ is the covariance matrix between ($\mu^i$, $x_1^i$, $c^i$). Hence, including the determinant $\mathrm{logdet} = \sum_i \log({\sigma_\mathrm{obs}^i}^2+\sigma_\mathrm{int}^2)$, we minimize $\chi^2 = \chi^2_\mathrm{res} + \chi^2_\mathrm{param} + \mathrm{logdet}$.

In our fit, we consider $p$ as non-noisy as it already takes into account errors on the environment parameter chosen to compute the magnitude step, so we assign it arbitrarily small errors $\sigma_p$, such that $p^\mathrm{true}$ is effectively equal to $p$. To avoid biases, the intrinsic scatter is fitted iteratively. We fix a given $\sigma_\mathrm{int}$, find the parameters ($\alpha$, $\beta$, $\gamma$, $M_0$, $(x_1^\mathrm{true})_i$, $(c^\mathrm{true})_i$) that minimise the total-$\chi^2$, then fix those parameters and fit $\sigma_\mathrm{int}$ as the value that normalises the reduced total-$\chi^2$. This procedure is repeated 10 times. The error estimation on the parameters is then done at fixed $\sigma_\mathrm{int}$ using a 1,000 steps MCMC. Finally, the corrected residuals are computed using the measured $c$, $x_1$ and $p$.

Based on realistic simulations, we show in Appendix~\ref{ap:fitting} that this fitting approach is unbiased, unlike the simple $\chi^2$ approach, which assumes $x_\mathrm{obs}=x_\mathrm{true}$. 
For a wide range of $\alpha$, $\beta$, $\gamma$ and $\sigma_\mathrm{int}$ combinations, we recover the input parameters and accurately estimate their errors, as our residual pulls (difference between the fitted and input parameter divided by the fitted error) are Gaussianly distributed, with a distribution centred on zero with a width of 1.

This fitting procedure makes use of \texttt{jax} \citep{jax2018github}. Additional details on the use of total-$\chi^2$ for cosmological parameter inference in SN cosmology will be given by Kuhn et al. (in prep).

\subsection{Standardisation parameter coefficients}
\label{sec:result_standardisation}

\begin{figure}
   \centering
   \includegraphics[width=0.8\columnwidth]{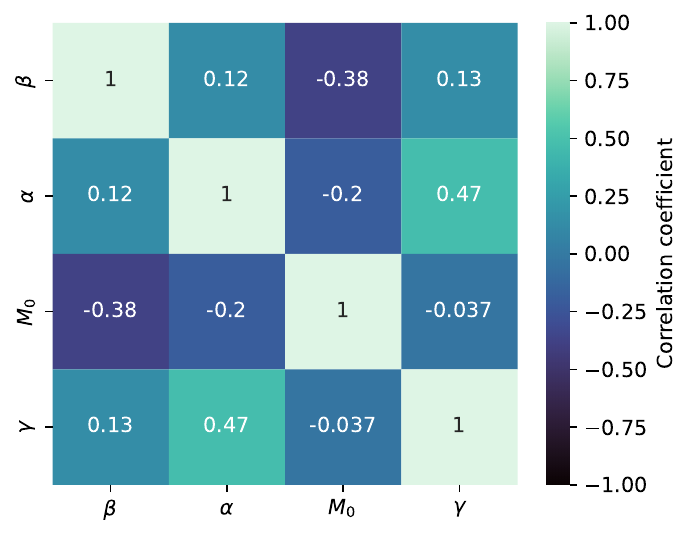}
      \caption{Correlation matrix of the standardisation parameters from Eq. \ref{eq:standardisation}.}
         \label{fig:corr_matrix_lin_alpha}
\end{figure}

\begin{figure}
   \centering
   \includegraphics[width=1\columnwidth]{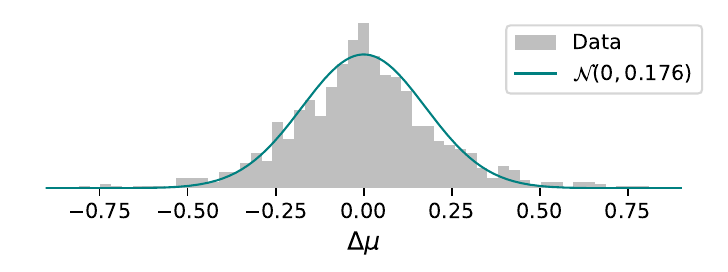}
      \caption{Histogram of the standardised Hubble residuals (corrected for SN colour, SN stretch and environment). The blue line is a Gaussian centred on 0 with a standard deviation corresponding to the normalised median absolute deviation $\sigma_\mathrm{nmad}=0.176$.}
         \label{fig:hubble_res}
\end{figure}

Fitting on our 938 SNe~Ia sample with the total-$\chi^2$ procedure described in Sect.~\ref{sec:totalchi2}, we find 
$\alpha=0.161\pm0.010$, $\beta=3.05\pm0.06$ and $\gamma=0.143\pm0.025$ using local colour as our environmental tracer \citep[e.g.,][]{Roman_2018, Kelsey_2021, Briday_2021}. $\sigma_\mathrm{int}$ is not released, as it will be the subject of a future detailed analysis.

The correlation matrix of the fitted ($\beta$, $\alpha$, $M_0$, $\gamma$) is plotted in Fig. \ref{fig:corr_matrix_lin_alpha}. We see a strong correlation between the stretch standardisation parameter $\alpha$ and the environmental step $\gamma$. This is expected, as $\gamma$ traces the magnitude bias due to environment and stretch is strongly linked with environment, as seen in Sect. \ref{sec:stretch_distribution}. 
Because of the correlation between $\alpha$, $\beta$ and $\gamma$, it is necessary to fit them jointly. Indeed, fitting for only $\alpha$ and $\beta$ and computing $\gamma$ as an a posteriori difference between the Hubble residuals for SNe in different environments will lead to $\alpha$ and $\beta$ absorbing part of the magnitude-environment connection. This will in turn bias $\alpha$ and $\beta$, and underestimate the step $\gamma$. This is discussed in more details in Sect. \ref{sec:steps}.

As illustrated in Fig.~\ref{fig:hubble_res}, the Hubble residuals are normally scattered, with a width (normalised median absolute deviation, nmad) of $0.175$ mag. The regular STD of the Hubble residuals is $0.226$ mag.
Discarding SNe Ia that have redshift extracted from SN Ia spectroscopic features (25\% of the sample, hence leaving 707/945 SNe~Ia) leads to very similar results ($\alpha=0.173\pm 0.009$, 
$\beta=3.06 \pm 0.05$ and $\gamma=0.143 \pm 0.022$), but with a reduced nmad of $0.16$ mag ($0.213$ for the regular STD). This reduction is expected since SN Ia-features redshift having a typical precision of $3\times10^{-3}$ (see details in \citealt{Smith_2024}), corresponding to an additional 0.08 mag scatter, to be added in quadrature.

The amplitude of the step $\gamma$ is discussed in Sect.~\ref{sec:steps}, $\beta$ is studied in a companion paper \citep{Ginolin_2024b} and $\alpha$ is further discussed in the following subsections.

\subsection{Environmental dependency of the stretch standardisation}
\label{sec:alpha_env}

The universality of the brighter-slower and brighter-bluer empirical linear relations might be challenged by observed environmental dependencies of such standardised SN~Ia magnitudes, e.g. the mass-step \citep[e.g.][]{Sullivan_2010}, the local colour bias \citep[e.g.][]{Roman_2018} or the age bias \citep{Rigault_2020}, that most likely are different aspects of the same underlying effect \cite[e.g.][]{Briday_2021,Brout_Scolnic_2021}.

In this subsection, we thus test if the stretch standardisation coefficient $\alpha$ itself is environment dependent. Using the environmental tracers introduced in Sect.~\ref{sec:environmental_properties}, we split our volume-limited sample in two, following the cuts introduced in that section (see also Table~\ref{tab:alpham0_env}). For each of these four environmental tracers, we independently standardise the two SN subsamples. We report in Table~\ref{tab:alpham0_env} the recovered stretch ($\alpha$) and absolute magnitude ($M_0$) parameters. The colour term ($\beta$) is discussed in a dedicated paper \citep{Ginolin_2024b}. The $M_0$ offset, equivalent to the step, is discussed in Sect.~\ref{sec:steps}.

\begin{table}
\centering
\tiny
\caption{Stretch correction coefficient ($\alpha$) and absolute magnitude ($M_0$) as a function of the SN environment.}
\label{tab:alpham0_env}
\begin{tabular}{l  l  c  c } 
\hline\\[-0.8em]
\hline\\[-0.5em]
Tracer & Cut &$\Delta\,\alpha$ & $\Delta\,M_0$\\
\hline\\[-0.5em]
$(g-z)_\mathrm{local}$ & 1 mag & $-0.032\pm0.021$ (1.5$\sigma$) & $0.123\pm0.021$ ($5.9\sigma)$\\[0.30em]
$(g-z)_\mathrm{global}$ & 1 mag & $-0.022\pm0.020$ (1.1$\sigma$) & $0.112\pm0.022$ ($5.0\sigma)$\\[0.30em]
$\log(M_\star/M_\odot)_\mathrm{local}$& 8.9 & $-0.041\pm0.019$ (2.2$\sigma$) & $0.128\pm0.022$ ($5.9\sigma)$ \\[0.30em]
$\log(M_\star/M_\odot)_\mathrm{global}$& 10 & $-0.059\pm0.020$ (3.0$\sigma$) & $0.135\pm0.021$ ($6.3\sigma)$\\[0.30em]
\hline
\end{tabular}
\end{table}

We see in Table~\ref{tab:alpham0_env} that the stretch standardisation parameter $\alpha$ does not seem to strongly depend on the environment. The most significant difference appears when using global host mass as tracer, with $\Delta \alpha \sim -0.06$ at the $3.0\sigma$ level, such that high-mass galaxies host SNe~Ia with a larger $\alpha$ coefficient.  This is consistent with earlier findings that the standardisation coefficients do not seem to depend on environment given the small data sets these analyses had access to \citep[e.g.][]{Sullivan_2010,Rigault_2020,Kelsey_2021}. The small dependence of $\alpha$ on environment is actually the signature of the non-linearity of the stretch-residuals relation, studied in detail in Sects. \ref{sec:alpha_linearity} and \ref{sec:broken_alpha_discussion}.

\subsection{Linearity of the stretch-magnitude relation}
\label{sec:alpha_linearity}

\begin{figure*}            
    \centering
     \includegraphics[width=1.6\columnwidth]{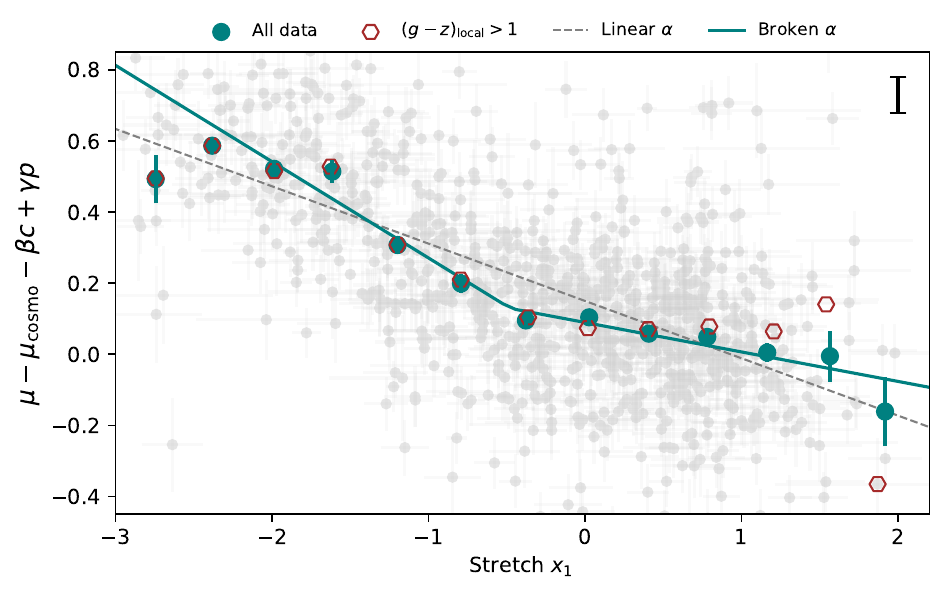}
   \caption{Standardised Hubble residuals ($\alpha=0$) as a function of stretch ($x_1$). 
      Large blue points show the data binned by stretch (mean) while the red hexagons show the binned stretch for SNe~Ia from locally red environments. The errorbars on the binned points only display the error on the mean, but a reference for the fitted intrinsic scatter $\sigma_\mathrm{int}$ is plotted in the top right corner. The blue line shows the best fitted broken-$\alpha$ model, while the dashed grey line shows the best $\alpha$ when assuming the linear Tripp relation.}
    \label{fig:broken_alpha}
\end{figure*}

The next standardisation assumption we challenge is the linearity of stretch-residuals relation \citep[see e.g. Figure~8 of][]{sullivan2011, Wang_2006, rubin2015}. More recently \cite{Garnavich_2023,  Larison_2024} presented $3$ to $4.5\sigma$ evidence for different  $\alpha$ coefficients as a function of SN stretch, using $\mathcal{O}(100)$ SNe~Ia.

We illustrate in Fig.~\ref{fig:broken_alpha} the stretch standardisation step for our sample. This figure shows the Hubble residuals standardised for colour and environment, but not for stretch, as a function of stretch ($x_1$). This relation was assumed to be linear by all SN cosmological analyses so far \citep[e.g.][]{Riess_1998,Perlumtter_1999,Betoule_2014,Scolnic_2018, Pantheonplus}. Yet, it is visible in Fig. \ref{fig:broken_alpha} that the stretch-residuals relation is not linear, as suggested by \cite{Garnavich_2023,  Larison_2024}. It indeed has a distinct broken shape, with lower-stretch SNe~Ia having a stronger (in absolute value) $\alpha$ coefficient than high-stretch SNe~Ia. Hints of this behaviour are also seen in the ZTF SN Ia DR2 siblings \citep{Dhawan_2024} and host-morphology \citep{Senzel_2024} studies.

To quantify this observation, we update Eq.~\ref{sec:standardisation} to allow for a broken-$\alpha$ standardisation. This introduces two extra parameters $\alpha_\mathrm{low}$ and $\alpha_\mathrm{high}$, in place of a single $\alpha$, and the breaking point $x_1^0$, which is also fitted:

\begin{equation}
\label{eq:brokerstandardisation}
      \mu_{\text{obs}} = m_B-M_0 - \beta c + \mathcal{A}(x_1)x_1 +p\gamma
\end{equation}
with
\begin{align}
    \label{eq:brokerstandardisation_alpha}
    \mathcal{A}(x_1) =
    \begin{cases}
    \alpha_\mathrm{low}  &\text{if $x_1 < x_1^0$}\\
    \alpha_\mathrm{high} &\text{otherwise}
    \end{cases}
\end{align}

Fitting this 'broken line' model, we find $\alpha_\mathrm{low}=0.271\pm0.011$ and $\alpha_\mathrm{high}=0.083\pm0.009$. The two $\alpha$ significantly differ at the $13.4\,\sigma$ level, with $\Delta \alpha = \alpha_\mathrm{high}-\alpha_\mathrm{low} = -0.188 \pm 0.014$. Such a high-significance demonstrates that indeed, the stretch-residuals relation is non-linear.
We also find a reduction of the fitted intrinsic scatter (-40\%). The reduction of the scatter indicates that the broken-$\alpha$ absorbs some of the unexplained scatter dubbed as intrinsic. While fitting for a broken-$\alpha$, we find $\beta_\mathrm{broken}=3.31 \pm 0.03$.
A detailed study of the intrinsic scatter will be the subject of a dedicated analysis and the actual $\sigma_\mathrm{int}$ values are thus not published.

The evaluation of the improvement of the fit by adding two extra parameters is not straightforward, as the $\chi^2$ of the fit depends on $\sigma_\mathrm{int}$, as explained in Sect. \ref{sec:standardisation}, which differs in the linear and broken-$\alpha$ case. If we fix $\sigma_\mathrm{int}$ to the broken-$\alpha$ value, we find $\Delta\mathrm{AIC}=101$. When assuming the larger $\sigma_\mathrm{int}$ associated to the linear-$\alpha$ model, we find $\Delta\mathrm{AIC}=48$, with the significance of the broken-$\alpha$ being reduced to $6.2\sigma$. In both cases, the broken-$\alpha$ model is favoured by the data.

We highlight that such a broken-$\alpha$ result is very unlikely to be caused by an unknown selection function. Indeed, this would cause to miss objects preferentially in the top-left corner of Fig. \ref{fig:broken_alpha} (faster and fainter objects), which would in turn pull $\alpha_\mathrm{high}$ to lower values. The robustness of this result is further demonstrated in Sect.~\ref{sec:tests}. 

The best-fit value of the stretch split $x_1^0$ is also of interest. When fitting the broken-$\alpha$ model with a variable breaking point, we find $x_1^0=-0.48\pm0.08$, which corresponds to the transition between the two stretch modes visible in the $x_1$ distribution in Fig.~\ref{fig:stretch_distrib}. This may suggest that there is a physical difference between the two stretch modes, resulting in the different stretch-residuals correlation. With the current modelling used, both modes share the same absolute standardised magnitude at the breaking point, hinting at a continuous transition between the two modes. This would need to be further investigated in a dedicated modelling analysis.
Additionally, we investigate in Sect. \ref{sec:broken_alpha_colour} the dependence of the broken-$\alpha$ on SN colour.

\subsection{Environmental dependency of the stretch-magnitude relation non-linearity}
\label{sec:alpha_linearity_and_env}

In Fig.~\ref{fig:broken_alpha}, we plot binned environment-standardised SN Hubble residuals as a function of stretch for locally red environment SNe~Ia only ($(g-z)_\mathrm{local} > 1$). Unlike the blue (younger) environments, these SNe~Ia populate both stretch modes (see Sect.~\ref{sec:stretch_distribution}), which enables us to test if indeed the broken relation is driven by the SN~Ia stretch mode (it should thus also be there for red-environment SNe~Ia) or if it is an artifact of an environmentally dependent $\alpha$ (the magnitude-stretch relation should then be linear). 
Considering only red-environment SN~Ia, we find $\Delta \alpha = -0.255 \pm 0.048$ ($5.3\sigma$) demonstrating that the broken-$\alpha$ effect is not caused by an environment-dependent $\alpha$, but is rather a stretch-mode driven effect, with each stretch modes having their own stretch-magnitude relation.
We reach the same conclusion when fitting a broken-$\alpha$ on SNe Ia in high-mass hosts, with $\Delta\alpha=-0.204\pm0.043$ ($4.8\sigma$).
Since the relative fraction of each mode is strongly environment-dependent  (cf. Sect.~\ref{sec:stretch_distribution}), this non-linearity leads to variation of the usual $\alpha$ standardisation as a function of the environment and redshift (see dedicated discussion in Sect.~\ref{sec:broken_alpha_discussion}).

\section{SN~Ia standardised magnitudes (steps)}
\label{sec:steps}

\begin{figure*}
   \centering
   \includegraphics[width=1.6\columnwidth]{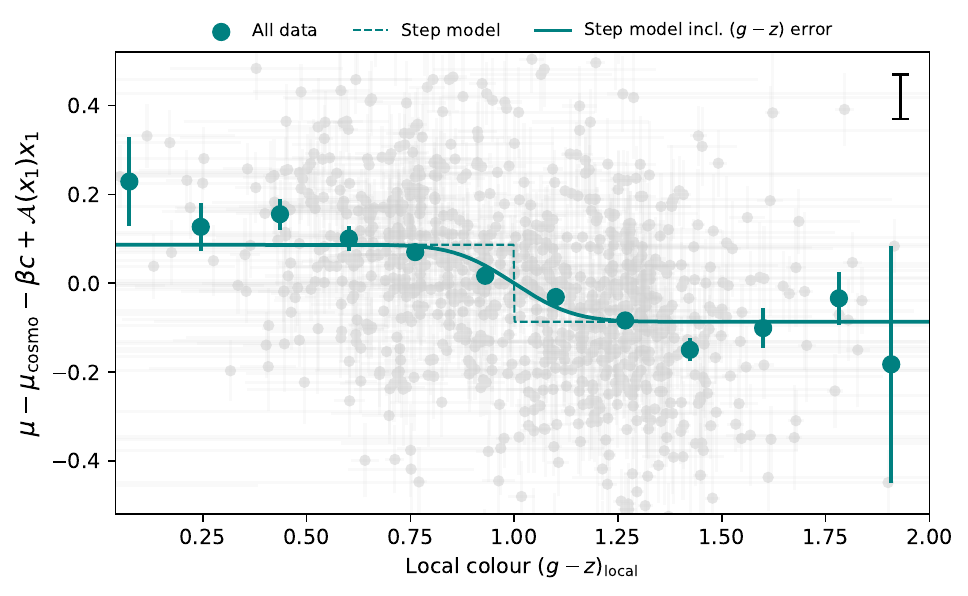}
      \caption{Standardised Hubble residuals corrected for stretch (with a broken-$\alpha$) and colour as a function of local environmental colour. Large blue points show the residuals binned by local colour, with the errorbars corresponding to the error on the mean. The scale of the fitted intrinsic scatter is shown in the top right corner. The dashed line shows the step model. The full line illustrates the step model once convolved by a Gaussian with a standard dispersion equals to the mean local environmental measurement error ($\sigma_{(g-z)\mathrm{, local}}\sim0.14$), as we account for those errors in the step model.}
         \label{fig:localcolour_step}
\end{figure*}

In this section, we investigate the amplitude of the environmental magnitude offsets affecting our volume-limited sample. 
These offsets between SN environments, aka steps, are represented by the $\gamma$ term in Eq.~\ref{eq:standardisation} and computed simultaneously with the other standardisation parameters.
We illustrate in Fig.~\ref{fig:localcolour_step} the local colour step, manually setting $\gamma=0$ to get Hubble residuals uncorrected for environment. This figure clearly demonstrates the existence of astrophysical biases affecting stretch and colour standardised SN~Ia magnitudes. Indeed, SNe~Ia in locally blue environments are significantly fainter ($\gamma=0.143 \pm 0.025$ mag) than those from locally red environments. Accounting for the non-linearity of the stretch-residuals relation, we find $\gamma=0.175 \pm 0.010$ mag.

Nonetheless, the actual origin of the observed astrophysical bias is still highly debated, and \cite{Brout_Scolnic_2021, Popovic_2021} suggest that the mass step originates from different dusts properties for different environments (see further discussion in \citealt{Kelsey_2023, Wiseman_2022}).
Modelling dust with BayeSN \citep{Thorp_2021, Mandel_2022}, \cite{Grayling_2024} find an intrinsic mass step of $-0.049\pm0.016$ mag. Finally, \cite{Smith_2020} shows that selection function corrections inaccuracy affects the ability to measure environmental biases, since all parameters are correlated (e.g. stretch, colour, host mass, local colour). The strength of our volume-limited data set is that we are free from such selection issues.

The fact that the steps, derived from a low-redshift volume-limited dataset, are found to be significantly larger than those from higher-redshift samples may be due to two things. First, the amplitude of the mass step may decrease with redshift \citep{Rigault_2013, Rigault_2020, Childress_2014}. Second, as the effect of selection function and the dependencies of SNe Ia with their astrophysical environment are intertwined, bias corrections, which are necessary at high-redshift, might affect the derivation of the step.
We note that the amplitude of our step is compatible with the value found by \cite{Rigault_2020} using another low-redshift, volume-limited sample (SNfactory, \citealt{Aldering_2002}).

The amplitude of the step is connected to the ability to accurately perform stretch and colour standardisation, since SN stretch (mostly) and colour are correlated with environmental parameters (see Fig.~\ref{fig:stretch_mass_lcolour} for the stretch-local colour connection). 
Hence, as already discussed in \citealt{Smith_2020, dixon2021, Briday_2021}, when fitting first for $\alpha$ and $\beta$, and then for $\gamma$ (as e.g. in \citealt{Kim_2019} for a recent example), the colour and stretch standardisation will absorb part of the (ignored) astrophysical biases, thus biasing all terms ($\alpha$, $\beta$ and $\gamma$ ; see, e.g. discussion in \citealt{Rigault_2020, Smith_2020, dixon2021, Briday_2021}). Such "a posteriori" measurements can then only underestimate the step.

\begin{figure}
   \centering
   \includegraphics[width=1\columnwidth]{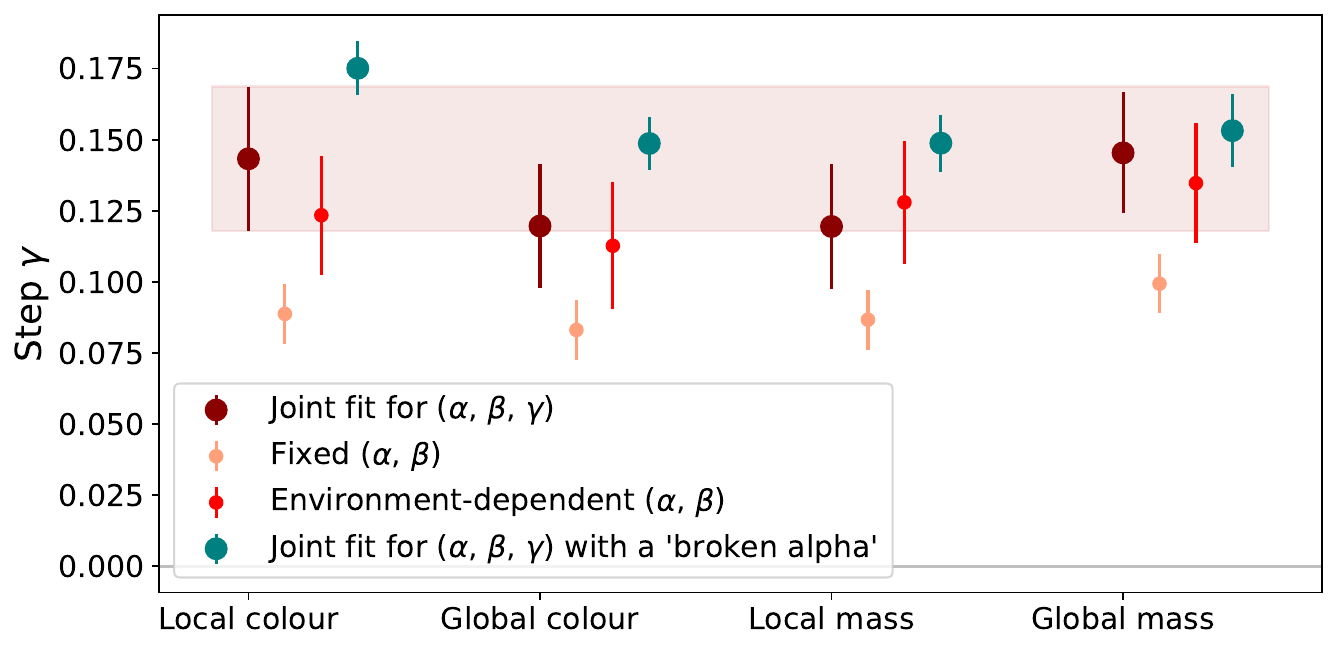}
      \caption{Fitted steps for each of the environmental proxies. The dark red and green points indicate steps where both ($\alpha$, $\beta$) and $\gamma$ are fitted at the same time for each proxy. The light red points indicate the steps when the same ($\alpha$, $\beta$) correction (fixed across proxies) is applied to the whole sample. The red points indicate the steps when independent ($\alpha$, $\beta$) corrections are applied for each of the subsamples. The green points are steps when including the broken-$\alpha$ model into the fit, according to Sect. \ref{sec:alpha_linearity}. The shaded band represents the $1\sigma$ interval for the reference step (local colour step from the joint fit).}
         \label{fig:steps_env}
\end{figure}

The best fitted $\gamma$ parameters are presented in Fig.~\ref{fig:steps_env} and reported in Table~\ref{tab:steps}. For comparison, the figure also includes the “a posteriori” step results, as well as the $\Delta M_0$ offset acquired when applying a per-environment standardisation and the steps obtained using a broken-$\alpha$ relation (see Sect.~\ref{sec:alpha_linearity}).

\begin{table}
\centering
\small
\caption{Amplitude of the environment magnitude offset, aka steps, for the joint fit for $(\alpha, \beta, \gamma)$, without and with broken-$\alpha$.}
\label{tab:steps}
\begin{tabular}{l c c } 
\hline\\[-0.8em]
\hline\\[-0.5em]
Tracer & $\gamma$ [mag] & $\gamma_{\mathrm{broken-}\alpha}$ [mag]\\
\hline\\[-0.5em]
$(g-z)_\mathrm{local}$ & $0.143 \pm 0.025$ ($5.6\,\sigma$) & $0.175 \pm 0.010$ ($18.4\,\sigma$)\\[0.30em]
$(g-z)_\mathrm{global}$ & $0.120 \pm 0.022$ ($5.5\,\sigma$) & $0.149 \pm 0.009$ ($16.0\,\sigma$)\\[0.30em]
$\log(M_\star/M_\odot)_\mathrm{local}$ & $0.120 \pm 0.022$ ($5.5\,\sigma$) & $0.149 \pm 0.010$ ($14.9\,\sigma$) \\[0.30em]
$\log(M_\star/M_\odot)_\mathrm{global}$ & $0.145 \pm 0.021$ ($6.8\,\sigma$) & $0.153 \pm 0.013$ ($12.0\,\sigma$) \\[0.30em]
\hline
\end{tabular}
\end{table}

All of magnitude offsets in Table \ref{tab:steps} are significantly non-zero at the $\geq 5\sigma$ level. When using the usual \cite{Tripp_1998} linear standardisation formalism, the strongest one is the global mass step with $\gamma= 0.145 \pm 0.021$ mag ($6.8\,\sigma$). The local colour step is at a similar level, in agreement with \cite{Roman_2018}, \cite{Kelsey_2021}, and \cite{Briday_2021}. The amplitude of the global mass step is consistent with findings reported in SNfactory \citep{Rigault_2020}, but higher than the steps reported in SDSS/SNLS \citep{Roman_2018} and DESY3 \citep{Kelsey_2021}.

As expected, Fig.~\ref{fig:steps_env} shows that the environmental steps derived after standardisation are significantly biased towards smaller values, with a reduction of $\sim0.05$ mag). 
When comparing the amplitude of magnitude offsets when the standardisation is made independently for the two environments, all steps remain strongly significant ($>5\,\sigma$ level), demonstrating that environmental inhomogeneity of the standardisation procedure is not responsible for SN magnitude biases, as already suggested in Sect.~\ref{sec:alpha_env}.
Furthermore, accounting for the non-linearity of the stretch-magnitude relation (cf. Sect.~\ref{sec:alpha_linearity}) slightly increases the amplitude of all the environmental steps and reduce the parameter errors, see Fig.~\ref{fig:steps_env}. This is to be expected if indeed the non-linearity is true, since the fitted model is more representative of the data, and thus all parameters are better measured. This also demonstrates that the existence of two stretch modes having each their own $\alpha$ term is not responsible for the environmental magnitude offsets. Altogether, local tracers perform slightly better than global ones when using the broken-$\alpha$ standardisation.

To ensure that the discrepancies between the step values are indeed due to variations in the fitting procedure, we investigated the differences between these step fitting methods using simulations in Sect. \ref{sec:steps_sims}. We reproduce both the order of the points and the factor between the jointly-fitted step (red points) and the a-posteriori step (light red points) for the local colour step. This confirms that not performing a joint fit for the environmental step along with $\alpha$ and $\beta$ will underestimate it.

\section{Discussion}
\label{sec:discussion}

\subsection{Robustness tests}
\label{sec:tests}

In this section, we test the robustness of our main results, namely the evolution of the stretch modes with global mass ($K_M$ parameter from Eq. \ref{eq:stretch2D_3}), the non-linearity of the stretch-residuals relation and the amplitude of the magnitude steps, by applying several analysis variations:
\begin{itemize}
    \item Varying the redshift cut we use to define our volume limited sample, reducing it to $z_\mathrm{max}=0.05$ as a conservative cut to test if our results could be caused by leftover selection function. We also extend our analyses to $z_\mathrm{max}=0.07$, to gain in statistical power, with a limited Malmquist bias.

    \item Only using SNe Ia with host redshifts, whose residual scatter is smaller.
    \item Changing the SN~Ia subpopulations used in the study. We include the subluminous 91bg population as well as SNe Ia classified as peculiar, as these could be hard to identify in higher redshift surveys for instance, or in surveys relying on photo-typing. We also redo the analysis discarding 91t. This allows us to test that our results are not driven by the 91t subpopulation.

    \item Using the literature cut on SN colour ($c<0.3$), to make sure our results are not affected by the inclusion of red objects.
    \item Applying a much stronger cut on lightcurve fit probability ($\chi^2_\texttt{SALT} < 0.1$), to check if our results could be induced by ill-modelled lightcurves.

\end{itemize}
The result of all these variations are summarised in Table~\ref{tab:tests}.

\begin{table}
\centering
\tiny
\caption{Robustness tests as described in Sect. \ref{sec:tests}.}
\begin{tabular}{l c c c c c} 
\hline\\[-0.8em]
\hline\\[-0.5em]
Test & $N_\mathrm{SN}$ & $K_M$ & $\Delta\alpha$ & $\gamma$ & $\gamma_{\mathrm{broken-}\alpha}$ \\
    & & [$\sigma$] & [$\sigma$] & [mag] & [mag]\\
\hline\\[-0.5em]
Fiducial & 945 & 9.2 & 13.4 & $0.143\pm0.025$ & $0.175\pm0.010$\\
\hline\\[-0.5em]
$z_\mathrm{max}=0.05 $ & 613 & 7.9 & 11.9 & $0.164\pm0.028$ & $0.204\pm0.011$\\[0.30em]
$z_\mathrm{max}=0.07 $ & 1341 & 10.6 & 14.7 & $0.143\pm0.017$ & $0.182\pm0.007$\\[0.30em]
Host $z$ & 707 & 6.9 & 9.7 & $0.143\pm0.024$ & $0.163\pm0.011$\\[0.30em]
incl. Ia-pec & 965 & 9.7 & 13.7 & $0.145\pm0.022$ & $0.182\pm0.010$\\[0.30em]
no 91t & 878 & 8.8 & 15.0 & $0.150\pm0.023$ & $0.172\pm0.012$ \\[0.30em]
$c<0.3$ & 852 & 9.1 & 10.0 & $0.144\pm0.022$ & $0.167\pm0.012$\\[0.30em]
$\chi^2_{\texttt{SALT}} < 0.1$ & 735 & 8.4 & 12.9 & $0.148\pm0.021$ & $0.182\pm0.011$\\[0.30em]
\hline\\[-0.5em]
Before Oct. 2019 & 451 & 5.4 & 7.3 & $0.132\pm0.032$ & $0.170\pm0.011$\\[0.30em]
After Dec. 2019 & 420 & 6.8 & 11.8 & $0.162\pm0.033$ & $0.201\pm0.013$\\[0.30em]
\hline
\end{tabular}
\vspace{5pt}
\tablefoot{Number of SNe in the sample, significance of the evolution of the stretch modes with colour ($K_M$ parameter from Eq. \ref{eq:stretch2D_3}), difference in $\alpha$ when fitting a broken standardisation law (using a variable cut) and step in local $(g-z)$ (with and without broken-$\alpha$ standardisation), for each of the tests performed in Sects. \ref{sec:tests}-\ref{sec:pocket_effect}.}
\label{tab:tests}
\end{table}

We see no significant variations across all modifications made. Accessing an even more secured volume limited sample only slightly reduces the significance of our results, as we reduce our statistics, strengthening our claim that the broken-$\alpha$ or the step amplitudes are not caused by any uncontrolled selection effect leftover. Going up to $z_\mathrm{max}=0.07$ does not changes our results, which might hint that the volume-limited cut we made was conservative. Only using host redshifts reduces the significance of the evolution of stretch with global mass. This is expected, as the host redshift sample is biased towards higher mass hosts (cf. Sect. \ref{fig:stretch_env_origin}), thus reducing the leverage of the fit. The significance of the broken-$\alpha$ also decreases, due to the reduced statistics.
Adding the peculiar 91bg-like objects, discarding 91t from our sample or only using $c<0.3$ SNe does not impact the significance of our results. The significance of the broken-$\alpha$ decreases slightly when discarding $c>0.3$ objects, but is still higher than the $10\sigma$ level. This confirms that red SNe do not exhibit peculiar behaviour, confirming the conclusions of \cite{Rose_2022} and \cite{Ginolin_2024b}. Enforcing a strong lightcurve fit requirement also has only a slight impact on $K_M$ and $\Delta\alpha$, due to lower statistics, confirming that it is not a lightcurve modelling issue that may cause either the broken stretch-residuals relation or the observed astrophysical bias (see a dedicated lightcurve residual analysis in \cite{Rigault_2024b}. 

\subsection{Dependence of the broken-$\alpha$ on SN colour}
\label{sec:broken_alpha_colour}

\begin{figure}
   \centering
   \includegraphics[width=1\columnwidth]{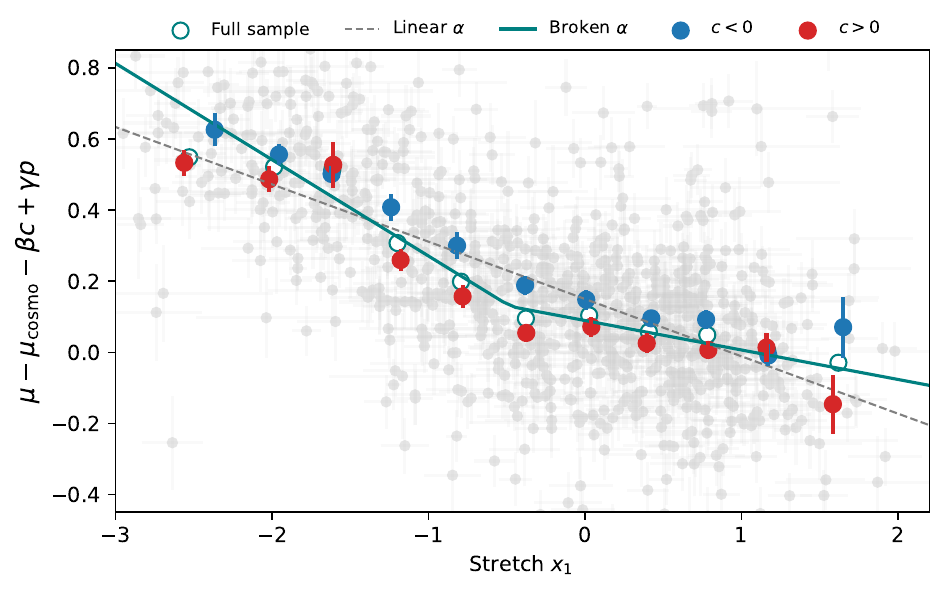}
      \caption{Hubble residuals corrected for colour and environment as a function of stretch $x_1$ (same as Fig. \ref{fig:broken_alpha}), for the full sample (white points) and blue/red SNe (blue/red points).}
         \label{fig:broken_alpha_colour}
\end{figure}

As an additional test of the broken-$\alpha$, we looked at its value when cutting the sample into red ($c>0$, 559 SNe) and blue ($c<0$, 379 SNe) SNe. We show in Fig. \ref{fig:broken_alpha_colour} the residuals corrected for colour and environment against stretch. The broken-$\alpha$ is still visible for both subsamples. We find $\Delta\alpha$ to be respectively $8.5\sigma$/$5.1\sigma$ away from 0 for the red/blue subsample. The reduction of the significance is expected due to the lower number of SNe in each subsample. We thus conclude that the non-linearity of the stretch-residuals relation does not depend on SN colour.

We note in Fig. \ref{fig:broken_alpha_colour}  that the blue SNe Ia residuals are $\sim0.07$ mag fainter than the red SNe Ia ones. As detailed in Appendix A of \cite{Ginolin_2024b}, this offset is caused by measurement noise in colour, that creates an up-tail towards blue values on correctly standardised Hubble residuals. We highlight that this is not a signature of a non-linearity in $\beta$; see Sect. 4.1 of \cite{Ginolin_2024b}, which confirms that the colour-residuals relation is linear.

\subsection{Pocket effect}
\label{sec:pocket_effect}

As discussed in Sect. \ref{sec:data}, the "pocket effect" that appeared in some of the ZTF CCDs after October 2019 can impact stretch measurements (Lacroix et al. (in prep), \citealt{DR2_overview}), leading to an overall shift in stretch of $\Delta x_1\sim-0.1$ for these SNe Ia.

In this section, we investigate the impact of this effect on our results by cutting our sample in two: SNe with $t_0$ before the 1st of October 2019 (451 SNe Ia), thus unaffected by the pocket effect, and those with $t_0$ after December 2019 (420 SNe Ia). SNe Ia at peak in October and November are discarded, as their lightcurves are only partially affected.

We first check that our data is indeed impacted by a stretch shift. We thus refit the double Gaussian stretch model presented in Sect. \ref{sec:stretch_1D}, but allowing for a shift between the two samples.
The recovered parameters of the double Gaussian are similar to those from the full sample (within $1\sigma$) and the fitted shift is $\Delta x_1 = -0.1\pm0.06$, in agreement with simulations.

To check that our main results are unaffected by the pocket effect, we then report in Table \ref{tab:tests} the values of the evolution of the stretch modes with host mass, the significance of the broken-$\alpha$ and the environmental steps for SNe Ia before/after the pocket effect, as was done in Sect. \ref{sec:tests}.

We first look at the dependency of the stretch mode means with the host stellar mass, parametrised by $K_M$. The significance of $K_M$ reduces from $9\sigma$ (full sample) to $\sim6\sigma$ for SNe~Ia from both subsamples. Such a lower significance is expected given the reduction of the sample size. Indeed, randomly drawing 450 SNe~Ia from the full sample and refitting for $K_M$ leads to a $6\sigma$ or lower signal 36\% of the time. Moreover, the values of $K_M$ are similar between subsamples ($K_M=-0.33\pm0.05$ for SNe Ia after Dec. 2019 and $K_M=-0.24\pm0.04$ for SNe Ia before Oct. 2019), and similar to the value for the full sample ($K_M=-0.30\pm0.03$).

The amplitude of the broken-$\alpha$ shows the strongest variation between the two samples: $7.3\sigma$ for the 451 SNe~Ia unaffected by the pocket effect and $11.8\sigma$ for the 420 SNe Ia impacted by it. It thus raises the question if this difference is simple causes for a reduction of the sample size. To evaluate this, we randomly select 450 SNe Ia from the full sample and fit for a broken-$\alpha$. Doing this process 1,000 times, we find $22\%$ of the downsized sample have $\Delta\alpha<7.3\sigma$ (96\% of $11.8\sigma$ of less).
Furthermore, the values of $\Delta\alpha$ are compatible between subsamples: for the SNe Ia before Oct. 2019, we find $\alpha_\mathrm{low}=0.257\pm0.014$ and $\alpha_\mathrm{high}=0.131\pm0.010$, while for the SNe Ia after Dec. 2019 we find $\alpha_\mathrm{low}=0.300\pm0.014$ and $\alpha_\mathrm{high}=0.086\pm0.011$. 
We thus conclude that the pocket effect does not significantly impact the broken-$\alpha$.

We finally investigate the environmental step amplitudes. 
The steps are similar between samples (less than a $\sim1.5\sigma$ difference), both for the linear and broken-$\alpha$ standardisation. We thus conclude that the pocket effect has no significant impact on the amplitude of the steps reported in this paper.

\subsection{Broken $\alpha$ and linear standardisation biases}
\label{sec:broken_alpha_discussion}

As presented in Sect.~\ref{sec:alpha_linearity}, the \cite{Phillips_1993} stretch-residuals relation, parametrised by the $\alpha$ term in Eq.~\ref{eq:standardisation}, is non-linear. This non-linearity coincides with the lightcurve stretch bimodality (cf. Fig.~\ref{fig:stretch_distrib} and, e.g. \citetalias{Nicolas_2021}), such that high-stretch mode SNe~Ia have a weaker stretch-residuals correlation ($\alpha_\mathrm{high}\sim0.08$) than those from the low-stretch mode ($\alpha_\mathrm{low}\sim0.27$, see Fig.~\ref{fig:stretch_distrib} and dedicated discussion in Sect.~\ref{sec:alpha_linearity}). Yet, the low-stretch mode is only accessible to SNe~Ia from old population environments
(see Sect.~\ref{sec:stretch_1D_env} and e.g. \citealt{howell2007}, \citealt{Sullivan_2010}, \citealt{Rigault_2020}, \citealt{Nicolas_2021}), unlike the high-stretch mode, which is always available. 
Since the fraction of old-population SNe~Ia varies with redshift and/or with environment, the use of a (usual) linear standardisation parameter $\alpha$ will lead to variations of $\alpha$ with redshift and environment. This is particularly important in two cases, redshift evolution and selection bias.

We here evaluate qualitatively the effect of fitting a linear $\alpha$ if the truth is a broken $\alpha$. A more thorough investigation of this problem, along with simulations, will be presented in a future paper.

Following \cite{Rigault_2020}, the fraction of SNe Ia from young progenitors increases with redshift, going to almost 1 at $z\sim2$. As they only populate the high-stretch mode, which has $\alpha_\mathrm{high}\sim0.08$, a linear $\alpha$ will evolve from its main sample value ($\alpha\sim0.16$) to a value close to $\alpha_\mathrm{high}$ when going to higher redshifts. We thus predict a decrease of a linear $\alpha$ with redshift.

Following the same reasoning, as low-strech SN Ia are fainter, lowering the magnitude limit of a survey thus increases the fraction of SNe Ia in the high-stretch mode, and biases a linear $\alpha$ towards $\alpha_\mathrm{high}$. We thus predict a decrease of a linear $\alpha$ when discarding objects due to a magnitude limit. 
Additionally, as shown in Sect. \ref{sec:stretch_distribution}, stretch is strongly correlated with environment. Indeed, SNe Ia in locally blue environments and/or low mass hosts only have access to the high-stretch mode. Hence, selecting SNe Ia only in those environments would bias a linear $\alpha$ towards $\alpha_\mathrm{high}$ (and towards $\alpha_\mathrm{low}$ for locally red environments and/or high mass hosts). Yet, it is common for SN Ia samples to be biased towards a given environment, e.g. mid-mass blue spiral galaxies, as it is easier to get a host redshift, massive galaxies, as it is easier to get a host mass, or targeted surveys that only observe massive spirals, to maximise the acquired SNe Ia rate.
The effect of magnitude and environment are likely to add up, making disentangling their contributions even harder. This thus calls for caution when fitting a linear stretch-residuals standardisation, especially on non volume-limited surveys. A more qualitative study of these effects, as well as the impact on the resulting fitted cosmology, will be presented in a upcoming analysis.

\section{Conclusion}
\label{sec:conclusion}

This paper presents a detailed analysis of the SN~Ia standardisation procedure, focusing on the stretch ($x_1$) residuals relation and SN environment magnitude offsets, aka steps. The companion paper \cite{Ginolin_2024b} focuses on SN colour. 

This analysis is based on  the volume-limited data set ($z<0.06$) of $\sim1000$ SN~Ia from the SN Ia ZTF DR2 sample \citep{DR2_overview, Smith_2024}. This data set is free from significant selection function and therefore observed distributions, correlations and biases are representative of the true underlying SN~Ia nature.

Our conclusions are the following:
\begin{enumerate}
    \item The stretch-residuals relation, represented by the $\alpha$ term in the SN~Ia standardisation \citep{Tripp_1998}, is non-linear. Low-stretch SNe~Ia ($x_1\leq-0.48$) have a much stronger magnitude-stretch relation with $\alpha_\mathrm{low}=0.271\pm0.011$ than high-stretch mode SNe Ia $\alpha_\mathrm{high}=0.083\pm0.009$. This non-linearity is measured at the $13.4\sigma$ level.
    
    \item The environmental dependency of standardised SN~Ia magnitude is stronger than usually claimed. Accounting for the stretch-residuals non-linearity we find $\gamma=0.175\pm 0.010$ mag using the local (2 kpc) colour $(g-z)$ tracer, a $18.4\sigma$ detection. Even with a usual linear $\alpha$ \citep{Tripp_1998} relation, we find $\gamma=0.143\pm0.025$ mag (a $5.6\sigma$ detection). 
    
    \item The global mass step is at a similar level than the local environmental colour step, with $\gamma=0.145\pm0.021$ mag ($0.153\pm0.013$ mag, using a broken-$\alpha$).

    \item The stretch distribution is bimodal with mode parameters measured in close agreement with those derived by \cite{Nicolas_2021}. The low-stretch mode ($x_1\leq-0.5$) is only available to SNe~Ia from old-population environments (redder galaxies and/or more massive host), unlike the high-stretch mode.

    \item The stretch modes means decrease linearly as a function of global host stellar mass (at a $9.2\sigma$ significance). We show that this, in turn, causes an apparent evolution of the stretch mode with local colour. This change is likely caused by the progenitor metallicity, that more directly correlates with stellar mass than with local environmental colour.

    \item The complex stretch-environment connection can be summarised as follows: SNe~Ia have two stretch modes, each having their own $\alpha$. The low stretch mode is only available to SNe~Ia from old-population environments (i.e. to “delayed” SNe~Ia). The mean of the $x_1$ distribution slowly evolves as a function of host stellar mass, suggesting that SN metallicity influences $x_1$, such that higher metallicity (more massive host) leads to SNe~Ia with smaller stretch. 
    
    \item Because the stretch distribution evolves with redshift and as a function of environment, we expect that the \cite{Tripp_1998} linear $\alpha$ term should decrease as a function of redshift. This should also strongly vary as a function of selection effect, such as magnitude limit or those favouring for instance blue emission line galaxies and/or mass hosts.
    
\end{enumerate}

This analysis of the ZTF volume-limited sample has revealed unexpected SN~Ia behaviours, notably the strong non-linearity of the SN~Ia standardisation procedure. This would likely not have been possible without the combination of a large sample statistic and the fact that our sample is not significantly affected by any (potentially hard to model) selection function effects. As such, we are able to predict what such effects could cause on the derivation of the SN Ia standardisation parameters that, in turn, leads to the inference of cosmological parameters. Our well controlled sample also reveals large environmental SN magnitude offsets. 
Altogether, those astrophysical biases, induced by our limited understanding of what SNe~Ia truly are, are a dominating source of systematic uncertainties, along with photometric calibration. To mitigate them, we strongly recommend for future surveys like the Legacy Survey of Space and Time (LSST) and the Nancy Grace Roman Space Telescope to dedicate a fraction of their observing time to build a volume-limited SN~Ia data set covering as much of the SN parameter phase space as possible. This is particularly true for the redshift component as redshift dependent issues directly translate into biases on the derivation of cosmological parameters.

\begin{acknowledgements}
Based on observations obtained with the Samuel Oschin Telescope 48-inch and the 60-inch Telescope at the Palomar Observatory as part of the Zwicky Transient Facility project. ZTF is supported by the National Science Foundation under Grants No. AST-1440341 and AST-2034437 and a collaboration including current partners Caltech, IPAC, the Weizmann Institute of Science, the Oskar Klein Center at Stockholm University, the University of Maryland, Deutsches Elektronen-Synchrotron and Humboldt University, the TANGO Consortium of Taiwan, the University of Wisconsin at Milwaukee, Trinity College Dublin, Lawrence Livermore National Laboratories, IN2P3, University of Warwick, Ruhr University Bochum, Northwestern University and former partners the University of Washington, Los Alamos National Laboratories, and Lawrence Berkeley National Laboratories. Operations are conducted by COO, IPAC, and UW.
SED Machine is based upon work supported by the National Science Foundation under Grant No. 1106171
The ZTF forced-photometry service was funded under the Heising-Simons Foundation grant \#12540303 (PI: Graham).
This project has received funding from the European Research Council (ERC) under the European Union's Horizon 2020 research and innovation program (grant agreement n 759194 - USNAC). MMK acknowledges generous support from the David and Lucille Packard Foundation. UB, MD, GD, KM and JHT are supported by the H2020 European Research Council grant no. 758638. LG acknowledges financial support from the Spanish Ministerio de Ciencia e Innovaci\'on (MCIN) and the Agencia Estatal de Investigaci\'on (AEI) 10.13039/501100011033 under the PID2020-115253GA-I00 HOSTFLOWS project, from Centro Superior de Investigaciones Cient\'ificas (CSIC) under the PIE project 20215AT016 and the program Unidad de Excelencia Mar\'ia de Maeztu CEX2020-001058-M, and from the Departament de Recerca i Universitats de la Generalitat de Catalunya through the 2021-SGR-01270 grant. This work has been supported by the research project grant “Understanding the Dynamic Universe” funded by the Knut and Alice Wallenberg Foundation under Dnr KAW 2018.0067, {\em Vetenskapsr\aa det}, the Swedish Research Council, project 2020-03444 and the G.R.E.A.T research environment, project number 2016-06012. YLK has received funding from the Science and Technology Facilities Council [grant number ST/V000713/1]. This work has been supported by the Agence Nationale de la Recherche of the French government through the program ANR-21-CE31-0016-03. TEMB acknowledges financial support from the Spanish Ministerio de Ciencia e Innovaci\'on (MCIN), the Agencia Estatal de Investigaci\'on (AEI) 10.13039/501100011033, and the European Union Next Generation EU/PRTR funds under the 2021 Juan de la Cierva program FJC2021-047124-I and the PID2020-115253GA-I00 HOSTFLOWS project, from Centro Superior de Investigaciones Cient\'ificas (CSIC) under the PIE project 20215AT016, and the program Unidad de Excelencia Mar\'ia de Maeztu CEX2020-001058-M. LH is funded by the Irish Research Council under grant number GOIPG/2020/1387. SD acknowledges support from the Marie Curie Individual Fellowship under grant ID 890695 and a Junior Research Fellowship at Lucy Cavendish College.
  
\end{acknowledgements}

\bibliographystyle{aa}
\bibliography{Biblio}

\begin{appendix}

\section{Fitting procedure}
\label{ap:fitting}

As mentioned in Sect. \ref{sec:standardisation}, we use a "total-$\chi^2$ approach to fit for standardisation. This is motivated by the known bias introduced by the usual $\chi^2$ method when fitting for a line where both axis have measurement errors. 
This issue is described in \cite{Kelly_2007}, and detailed in \cite{Kowalski_2008} in a cosmological case (see also Appendix A of Ginolin et al 2024b for an example case on SNe standardisation). 

\begin{figure}
   \centering
   \includegraphics[width=1\columnwidth]{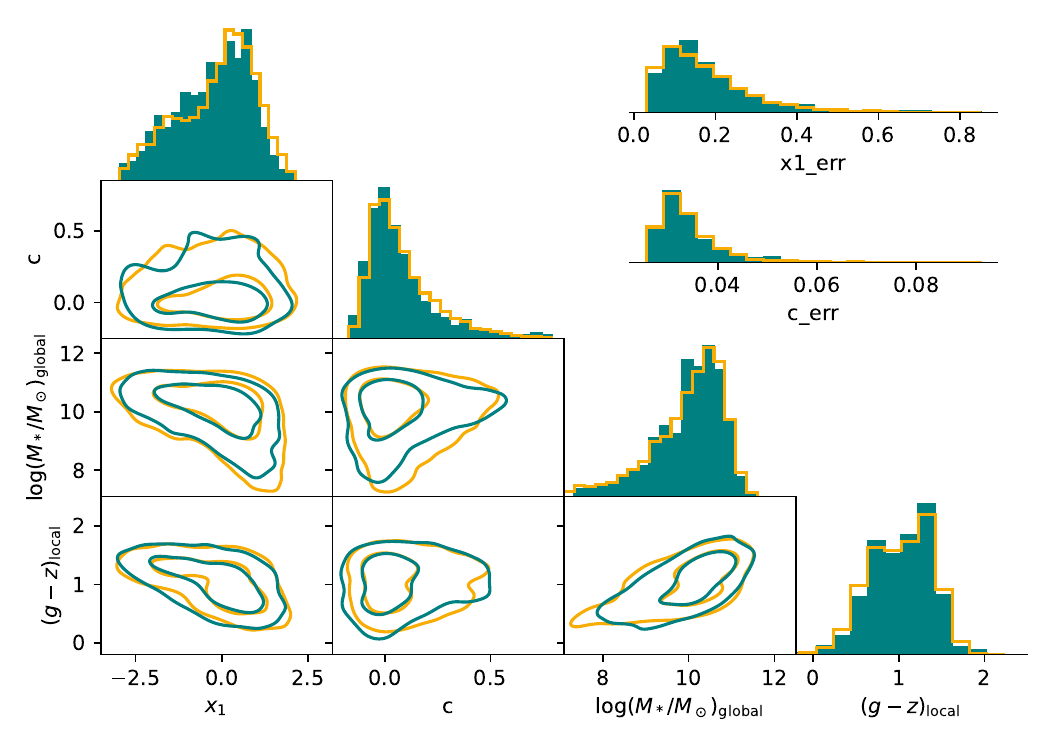}
    \caption{\textit{Corner plot:} Correlation between SN~Ia lightcurve stretch ($x_1$), colour ($c$), host mass and environmental local colour for the data (teal) and our realistic simulations (orange). Lines in the non-histogram panels show the contours containing 90\% and 50\% of the data (teal) and simulated points (orange).
    \textit{Top right:} Distribution of the stretch and colour errors following the corner-plot colour code.}
    \label{fig:simuskysurvey}
\end{figure}

In this section we present realistic simulations of our dataset on which we perform both the regular log-likelihood minimisation (simple $\chi^2$) and the total $\chi^2$ method used in this paper. These simulation are made using \texttt{skysuvey}\footnote{\url{skysurvey.readthedocs.io}}. We compare in Fig.~\ref{fig:simuskysurvey} our data with \texttt{skysurvey} simulations. In this analysis, to be able to quickly simulate variants of the survey, we bypass the lightcurve fitting step. We tested that for ZTF-like survey this method does not introduce any significant biases on the SN parameters or their errors. Both the errors and the correlations between lightcurves parameters and environment proxies are indeed realistic. A detailed study of the underlying SNe~Ia population using such realistic simulations is ongoing and will be the subject of a dedicated follow-up analysis.

\subsection{Total $\chi^2$ accuracy and simple $\chi^2$ biases}

Given this simulation setup, we generate a set of 81 (3x3x3x3) simulations forming a grid of all permutation of the following parameters: $\alpha \in [0.1, 0.15, 0.2]$, $\beta \in [2.5, 3.5, 4.5]$, $\gamma \in [0, 0.1, 0.2]$ and $\sigma_\mathrm{int} \in [0.07, 0.12, 0.17]$. Each time, we draw 2,000 SNe Ia and we simulate each permutation 5 times to get, at the end 405 simulations. For each of these simulation, we fit for $\alpha$, $\beta$, $\gamma$ and $\sigma_\mathrm{int}$ for both the simple $\chi^2$ and the total-$\chi^2$ methods. The regular $\chi^2$ minimisation assumes $x_\mathrm{obs}=x_\mathrm{true}$, while the total-$\chi^2$ method fits for $x_\mathrm{true}$ as nuisance parameters (cf. Sect. \ref{sec:totalchi2}).

\begin{figure}
   \centering
   \includegraphics[width=1\columnwidth]{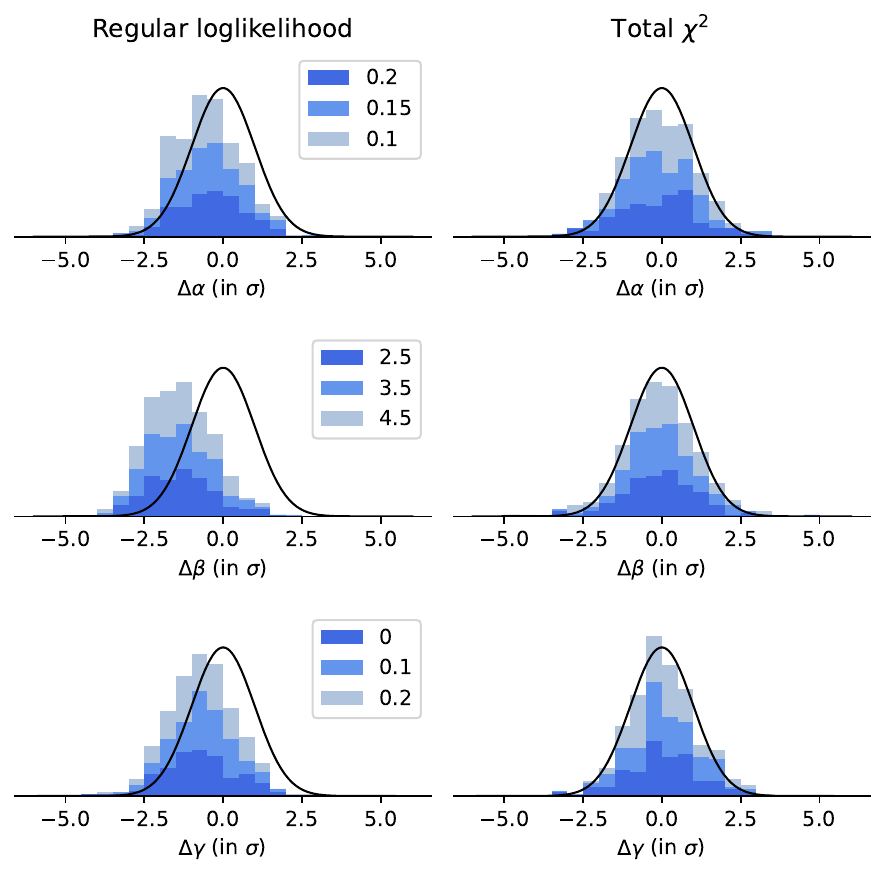}
    \caption{\textit{Left:} Difference between the fitted ($\alpha$, $\beta$, $\gamma$) and the input value when using a usual log-likelihood minimisation code, in $\sigma$. The stacked histograms corresponds to three different input values defined in the legend on the right. The black line is the model the histogram should follow if the results were unbiased (both values and error), which is a Gaussian with a null mean and unitary standard deviation. \textit{Right:} Same as the left plot, but for the total-$\chi^2$ method.}
    \label{fig:standardisation_test}
\end{figure}

We present in Fig.~\ref{fig:standardisation_test}, the results of these tests. The result are shown in the form of pulls, i.e. $(\mathrm{fit}-\mathrm{truth})/\mathrm{fit\_err}$. If the fit is accurate, the pull distribution should be centred on zero (unbiased) with the width of 1 (correct fit error). Figure \ref{fig:standardisation_test} clearly illustrate that the simple $\chi^2$ results are biased toward lower values (closer to zero) for all standardisation parameters, especially for $\beta$. This is because the typical measurement errors ($\overline{\sigma_c}=0.03$) are non-negligible in comparison to the typical underlying colour scatter ($\mathrm{STD}(c)=0.15$). On the opposite, standardisation parameters are always accurately recovered when using the total-$\chi^2$ used in this analysis and detailed in Section~\ref{sec:standardisation}.

\subsection{Significance of the broken magnitude-stretch relation}

We use our simulation tool to further test the veracity of the broken-$\alpha$ model. We simulate data using a single $\alpha$ and fit for a broken relation that split at the expected $x_1=-0.5$, and then compute the odds to find such a strong difference in $\alpha$ between the low and high-stretch modes.

\begin{figure}
   \centering
   \includegraphics[width=1\columnwidth]{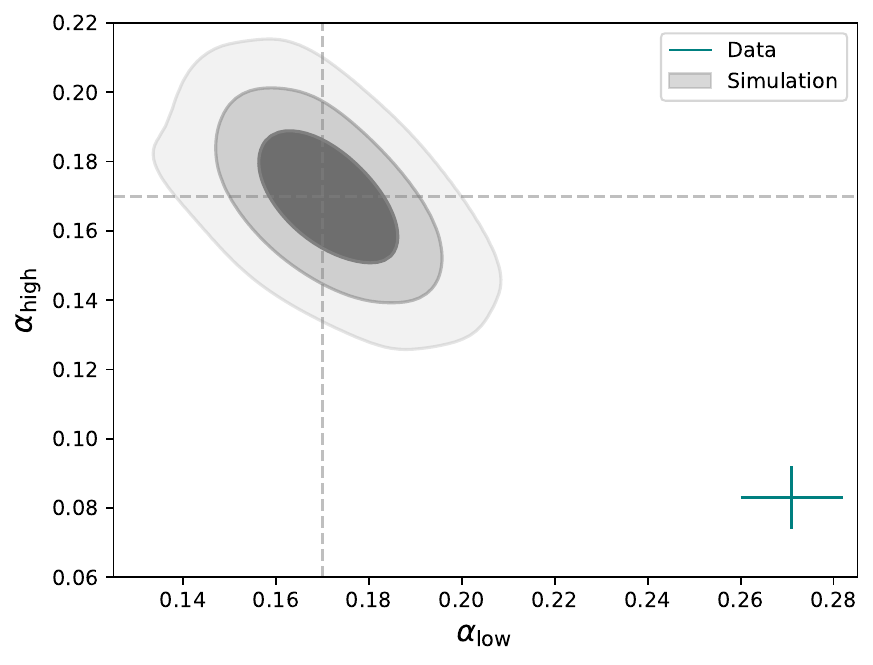}
    \caption{$\alpha_\mathrm{low}$ vs $\alpha_\mathrm{high}$ when fitting a broken-$\alpha$ on samples simulated with a linear $\alpha$. The dashed lines represent the input values. The contours (respectively 50\%, 84\% and 97\%) are computed with 1,000 simulations of a similar-sized sample than the one used throughout the paper, and the data point represent the values from the fit in Sect. \ref{sec:alpha_linearity}.}
    \label{fig:simu_broken_alpha}
\end{figure}

We thus simulate 1,000 samples of 927 SNe using a linear $\alpha$, setting $\alpha=0.17$, i.e. close to that obtain using our sample while fitting a single $\alpha$, and then perform the broken-$\alpha$ fit. 
We show in Fig.~\ref{fig:simu_broken_alpha} the resulting  distribution of the fitted $\alpha_{\mathrm{low}}$ and $\alpha_{\mathrm{high}}$. They are accurately recovered in close agreement with the input $\alpha_{\mathrm{low}}=\alpha_{\mathrm{high}}=0.17$. The $\alpha_{\mathrm{low}}=0.271\pm0.011$ and $\alpha_{\mathrm{high}}=0.083\pm0.009$ fitted on our data (teal point in Fig.~\ref{fig:simu_broken_alpha}) are thus confirmed to be significantly different, confirming the non-linearity of the residuals-stretch relation.

\subsection{Difference between the step fitting methods}
\label{sec:steps_sims}

\begin{figure}
   \centering
   \includegraphics[width=1\columnwidth]{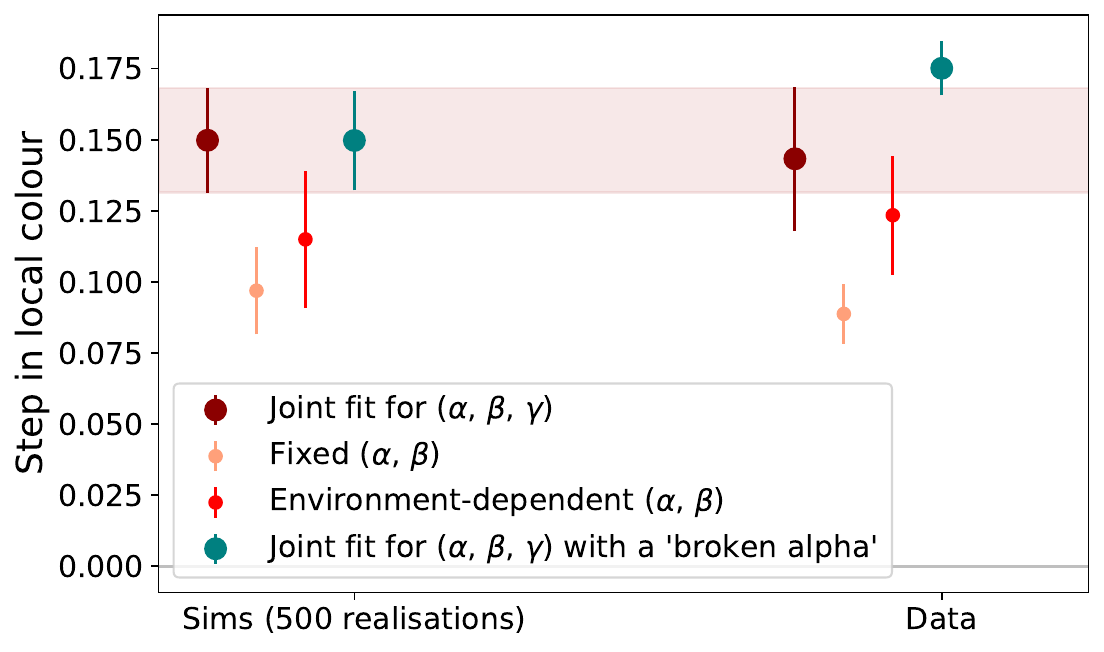}
    \caption{Local colour step for the four fitting procedures presented in Fig. \ref{fig:steps_env}, with 500 realisations of the simulations described in Sect. \ref{sec:steps_sims} (left) and data (right). The points for the simulations are the means of the realisations, and the errorbars are the standard deviation.}
    \label{fig:simu_steps}
\end{figure}

We investigated the step values obtained with the different methods described in Sect. \ref{sec:steps}, to validate that the discrepancies seen between the steps were due to variations in the fitting procedures. We used the simulations presented in Sect. \ref{ap:fitting}, with a broken-$\alpha$ and a local colour step of $\gamma=0.16$ mag. The results are presented in Fig. \ref{fig:simu_steps}. The order of the points seen in the data is well reproduced by the simulations. Indeed, the "Fixed ($\alpha$, $\beta$)" step is the smallest of the four steps, while the other three steps are all compatible. The factor between the regular step ("Joint fit for ($\alpha$, $\beta$, $\gamma$)") and the "Fixed ($\alpha$, $\beta$)" step is the same in the data and in the simulations. The data exhibits a step difference of $\Delta\gamma=0.055\pm0.027$ mag, while the simulations exhibits a $\Delta\gamma=0.052\pm0.024$ mag difference (a $0.05\sigma$ difference between data and simulations).
In the simulations, the broken-$\alpha$ step is of the same order as the regular linear-$\alpha$ step, while it is slightly higher in the data, with a reduced errorbar. However, we know the link between step and environment is more complex than the current model used in simulations, and that as environment strongly correlates with stretch, this will likely impact the broken-$\alpha$ step. A further modelling of an age step will be the subject of a future analysis.

\end{appendix}

\end{document}